\documentclass{aa}   

\usepackage{enumerate, graphicx}
\usepackage{adjustbox}
\usepackage{float}
\usepackage{txfonts}

\usepackage{hyperref}
\hypersetup{
    colorlinks=true,
    linkcolor=black,
    urlcolor=black,
    citecolor=blue,
    }
\usepackage{comment}

\usepackage{cancel}
\usepackage[shortlabels]{enumitem}
\usepackage{siunitx}
\DeclareSIUnit \h {\mbox{$h$}}
\DeclareSIUnit \parsec {pc}
\DeclareSIUnit \Msun {\mbox{M$_{\odot}$}}
\DeclareSIUnit \Lsun {\mbox{L$_{\odot}$}}
\DeclareSIUnit \century{century}
\DeclareSIUnit \year{yr}

\newcommand{\sex}{\textsc{SExtractor}}
\newcommand{\psfex}{\textsc{PSFEx}}

\newcommand{\dos}{{\sc SExtractor+PSFEx\ }} 
\usepackage{natbib}
\bibpunct{(}{)}{;}{a}{}{,} 
\begin{document}

\title{The VVV near-IR galaxy catalogue of the southern Galactic disc}  
\titlerunning{VVV near-IR Galaxy Catalogue III}

\author{M. V. Alonso\inst{1,2} \and L. D. Baravalle\inst{1,2} 
\and J. L. Nilo Castell\'on\inst{3} \and C. Villalon\inst{1,2} \and M. Soto\inst{4} \and M. A. Sgr\'o\inst{5,2} \and I. V. Daza Perilla\inst{6,7} 
\and C. Valotto\inst{1,2}  \and M. Lares\inst{1,2} \and D. Minniti\inst{8,9} \and P. Marchant Cortés\inst{3} \and F. Milla Castro\inst{3} \and M. Hempel\inst{8,10} \and J. Alonso-García\inst{11,12} \and L. Macri\inst{13} \and A. Pichel\inst{14} \and 
N. Masetti\inst{15,8} \and R. K. Saito\inst{16} \and M. G\'omez\inst{8} 
}
      
\institute{
Instituto de Astronom\'{\i}a Te\'orica y Experimental, (IATE-CONICET), Laprida 854, X5000BGR, C\'ordoba, Argentina.
\and
Observatorio Astron\'omico de C\'ordoba, Universidad Nacional de C\'ordoba, Laprida 854, X5000BGR, C\'ordoba, Argentina.
\and
Departamento Astronom\'ia, Universidad de La Serena. Av. 
Raúl Bitran 1305, La Serena, Región de Coquimbo, Chile. 
\and
 Instituto de Investigación en Astronomía y Ciencias Planetarias, Universidad de Atacama, Copayapu 485, Copiapó, Chile.
 \and
  Instituto de Altos Estudios Espaciales “Mario Gulich” (CONAE – UNC), Argentina.
  \and
 Center for Space Science and Technology, University of Maryland, Baltimore County, 1000 Hilltop Circle, Baltimore MD 21250, USA.
 \and
Center for Research and Exploration in Space Science and Technology, NASA/Goddard Space
Flight Center, Greenbelt, MD 20771, USA. 
\and
  Instituto de Astrof\'isica, Depto. de Ciencias F\'isicas, Facultad de Ciencias Exactas, Universidad Andr\'es Bello. Av. Fernandez Concha 700, Las Condes, Santiago, Chile.
 \and
 Vatican Observatory, V00120 Vatican City State, Italy.
\and 
 Max-Planck Institut for Astronomie, Königstuhl 17, 69117 Heidelberg, Germany
\and
Centro de Astronom\'{i}a (CITEVA), Universidad de Antofagasta, Av. Angamos 601, Antofagasta, Chile.
\and 
Millennium Institute of Astrophysics, Nuncio Monse\~nor Sotero Sanz 100, Of. 104, Providencia, Santiago, Chile.
\and
Department of Physics and Astronomy, The University of Texas Rio Grande Valley, 1201 W. University Drive, Edinburg, TX, 78539, USA. 
\and 
Instituto de Astronom\'ia y F\'isica del Espacio, CONICET–UBA, Av. Int. Guiraldes 2620, C1428BNB CABA, Argentina.
\and 
INAF - Osservatorio di Astrofisica e Scienza dello Spazio, via Piero Gobetti 101, I-40129 Bologna, Italy.
\and
Departamento de F\'isica, Universidade Federal de Santa Catarina, Trinidade 88040-900, Florianopolis, Brazil.
}

\date{Accepted XXX. Received YYY; in original form ZZZ}

\abstract
{The distribution of galaxies in the Zone of Avoidance (ZoA) is incomplete due to the presence of our own Galaxy. 
}
{Our research focused on the identification and characterisation of galaxies in the ZoA, 
using the new near-infrared (NIR) data from the VVVX survey in the regions that cover the southern Galactic disc (230$^{\circ}$ $<l<$ 350$^{\circ}$).
} 
{We used our previously-established procedure based on photometric and morphological criteria to identify galaxies. The large data volume collected by the VVVX required alternatives to visual inspection, including artificial intelligence techniques, such as classifiers based on neural networks.}
{The VVV NIR galaxy catalogue (VVV NIRGC III) is presented,  covering the southern Galactic disc, significantly extending the vision down to $K^0_s = 16$ mag throughout the ZoA. This catalogue provides positions, photometric and morphological parameters for a total of 167,559 galaxies with their probabilities determined by the CNN and XGBoost algorithms based on image and photometric data, respectively. The construction of the catalogue involves the employment of optimal probability criteria. 14\% of these galaxies were confirmed by visual inspection or by matching with previous catalogues. The peculiarities exhibited by distinct regions across the Galactic disc, along with the characteristics of the galaxies, are thoroughly examined.
The catalogue serves as a valuable resource for extragalactic studies within the ZoA, providing a crucial complement to the forthcoming radio catalogues and future surveys utilizing the Vera C.~Rubin Observatory and the Nancy Grace Roman Space Telescope.  
}
{We present a deep galaxy map, 
covering a 1080 sq. deg. region ($230^{\circ} \le l \le 350^{\circ}$ and  $|b| \le 4.5^{\circ}$), which reveals that the apparent galaxy density is predominantly influenced by foreground extinction from the Milky Way. However, the presence of intrinsic inhomogeneities, potentially associated with candidate galaxy groups or clusters and filaments, is also discernible.}
  {}

   \maketitle
%

\section{Introduction}\label{intro}

A fundamental requirement for our understanding of the local Universe and its dynamics is the availability of a comprehensive map of galaxies across the entire sky. However, a significant fraction of the sky (|b| $ < $ 15$^{\circ}$) is obscured by the Milky Way (MW), resulting in an area with a relatively limited number of observed galaxies. The presence of dust and foreground stars from our own Galaxy attenuates the light coming from extragalactic objects, making them fainter and smaller. This region is called the Zone of Avoidance (ZoA), as defined by \cite{Shapley1961}, and detecting galaxies is a major challenge. 
The ZoA covers a significant volume of the local Universe (z $<$ 0.05), and its study is essential for understanding nearby large-scale structures. This region also plays a critical role in determining the nature of the luminosity distances and redshifts, and in the interpretation of Hubble diagrams \citep{Clocchiatti2024}.

In recent years, considerable progress has been achieved in the ZoA region through the use of detectors of different wavelengths, including optical, infrared, radio, and X-rays. The specific wavelength employed in a survey determines the morphological type of galaxies that can be detected. For instance, near-infrared (NIR) radiation is more sensitive to early-type galaxies, whereas HI surveys are more conducive to the detection of late-type galaxies \citep{Williams2014}. 

Optical surveys are constrained by surface brightness, with only the most prominent and luminous galaxies discernible behind the Galactic plane \citep{Kraan1996,Woudt2001}. Infrared (IR) radiation can penetrate dust and gas, thus providing a valuable tool for extragalactic studies extending beyond our own Galaxy. In this sense, \cite{2000AJ....119.2498J} discussed the principal limitations to the detection of galaxies behind the Galactic plane, where the stellar density is higher, thereby rendering detection increasingly challenging. The Two Micron All Sky Survey (2MASS, \citealt{Skrutskie2006}) mapped the entire sky at NIR wavelengths, and yielded the 2MASS extended source catalogue (2MASX, \citealt{Jarrett2004}). The UKIDSS Galactic Plane Survey (GPS, \citealt{Lucas2008}) was another NIR survey that covered three ZoA  regions: 141$^{\circ} < l < $ 230$^{\circ}$, $|b| <  $ 5$^{\circ}$; 15$^{\circ} < l <  $ 107$^{\circ}$, $|b| <  $ 5$^{\circ}$; and -2$^{\circ}  < l < $ 15$^{\circ}$, $|b| <  $ 2$^{\circ}$. 

VISTA Variables in the Vía Láctea (VVV, \citealt{Minniti2010}) is a NIR survey specifically designed to probe deeper into regions of high interstellar extinction at low latitudes of the Galactic bulge and disc. The VVVX survey, an extension of the original VVV survey, encompasses three times the sky coverage in the ZoA \citep{Minniti2018, Saito2024}. This extension includes part of the northern Galactic disc (10$^{\circ}$ $<l<$ 20$^{\circ}$) and notably increases the area of the southern Galactic disc (230$^{\circ}$ $<l<$ 350$^{\circ}$). The principal objective of the project was to investigate the nature of variable stars in the MW and other Galactic objects, including open and globular clusters. In addition, this project aimed to gain insights into the structure of the Galaxy near its centre by utilising NIR microlensing techniques. The main goal of our project was to identify galaxies beyond our own Galaxy. Previous NIR surveys have not been able to reach these low Galactic latitudes. 

Using data from the VVV survey, several studies have been carried out to detect galaxies in the ZoA. \cite{Amores2012} conducted a visual identification of 204 galaxy candidates in the region around $l =298^{\circ}.4$, $b = -1^{\circ}$, utilising a combination of their size and colours as identifying criteria. In the Galactic bulge region, \cite{Coldwell2014} and \cite{Galdeano2021, Galdeano2022,Galdeano2023} conducted studies on the NIR properties of different galaxy clusters. Recently, \cite{Duplancic2024} reported the discovery of over 14,000 new galaxy candidates in that same region.

In a series of papers, we employed a procedure based on  \sex\ \citep{Bertin1996} and \psfex\ \citep{Bertin2011} using the VVV and VVVX images to identify and characterise galaxies behind the Galactic disc. The selection was based on morphological parameters and colours, and we subsequently confirmed the galaxies through visual inspection. The initial study in the series, \cite{Baravalle2018}, established the methodology and criteria for selecting galaxies, with visual inspection being a crucial step in this phase of the process. \cite{Baravalle2021} presented the VVV NIR Galaxy Catalogue (VVV NIRGC), the first catalogue of galaxies in the southern Galactic disc based on the selection of candidates, with visual inspection carried out on all of them to discard false identifications. This catalogue will be referred to as VVV NIRGC I and includes 5,554 galaxies, of which only 45 were previously known.

Working with the VVVX data in parts of the northern and southern Galactic disc involves about four times the area covered by the VVV survey. The application of machine learning techniques to this vast amount of data allows us to discriminate between genuine galaxies and false identifications. In \cite{Daza2023}, we devised a procedure capable of performing an automatic identification of galaxies and non-galaxies in the VVV and VVVX surveys, which we applied to images covering the northern Galactic disc. The objects were identified and characterised with {\sc SExtractor+PSFEx}, and the selection of galaxy candidates was based on morphological parameters and colours, as used previously. Then, statistical methodologies were employed, including unsupervised and supervised machine learning techniques, which were implemented on images and photometric information independently. 
The resulting catalogue, comprising 1,003 galaxies, which had one of the convolutional neural network (CNN, \citealt{Cheng_2021}) or the gradient boosting model (XGBoost, \citealt{Chen2016}) probabilities  $\geq$ 0.6 based on the results of machine learning algorithms, will be hereafter referred to as VVV NIRGC II. To validate this
procedure, all the galaxies were confirmed by visual inspection. In this region, there were only two previously-known galaxies.

The 21-cm HI emission is also capable of penetrating the ZoA without being hindered by dust, hence allowing the observation of otherwise obscured regions \citep{Henning1992}. The main result of many systematic HI surveys in the regions of the Galactic disc (e.g., \citealt{Ramatsoku2016,Staveley2016,Kraan2018}) was to measure radial velocities of gas-rich galaxies, which allows us to map large-scale structures and improve our knowledge of the distribution of matter in regions that cannot be accessed at optical {or NIR wavelengths.  
Recently, the HI data from the SARAO MeerKAT Galactic Plane survey \citep{Goedhart2024} enabled the detection of 477 candidate galaxies tracing the Great Attractor wall \citep{Steyn2024} and 1,562 galaxies in the region of the Vela Supercluster \citep{Rajohnson2024}. 

This paper presents the final contribution to the series to find and search for galaxies behind the Galactic disc. It is organised as follows: In \S\ref{data}, we provide a brief overview of the VVV and VVVX surveys. In \S\ref{procedure}, we present the methodology used to find galaxies at low Galactic latitudes.  \S\ref{sec:catalogue}  presents the galaxy catalogue (VVV NIRGC III), as well as estimates of the internal photometric errors and comparisons with other surveys. In \S\ref{sec:analysis} and \ref{sec:distribution}, we analyse the photometric properties and the distribution of galaxies, and in \S\ref{sec:conclusions}, we present a summary of the work and make suggestions about the next steps in these ZoA regions.
We use the following cosmological
parameters: H$_{0}$ = 70.4 kms$^{-1}$ Mpc$^{-1}$,
$\Omega_{M}$ = 0.272, and $\Omega_{\lambda}$ = 0.728 
\citep{Komatsu2011}.


\section{The VVV and VVVX surveys: the Galactic disc data}\label{data}

The VVV survey \citep{Minniti2010} probed a section of the southern Galactic disc using five near-IR passbands: $Z$, $Y$, $J$, $H$ and $K_{s}$. The VVVX survey \citep{Minniti2018, Saito2024} extended the coverage into the northern and southern Galactic discs using only three passbands: $J$, $H$, and $K_{s}$. Both surveys were divided into tiles of 1$^{\circ}$  $\times$ 1.5$^{\circ}$, which were generated by six single-point observations. The primary goal of the surveys was to study the variability of stars in our Galaxy. However, we also used the NIR data to identify galaxies situated behind the Galactic disc.    

Table~\ref{tab:table0} shows the different regions of the Galactic disc covered by the VVV and VVVX surveys and the VVV NIRGC catalogues. The VVV survey has been used to study the inner parts of the southern Galactic disc, comprising 152 tiles in the disc region. 
The VVV NIRGC I catalogue \citep{Baravalle2021} contains the photometric information of 
5,554 galaxies.  
Furthermore, \cite{Daza2023} extended the search for galaxies using the VVVX survey of the northern Galactic disc, which includes 56 tiles in the disc+20 region, 
with the VVV NIRGC II catalogue containing 1,003 galaxies. 
In the current paper, we take advantage of the 512 VVVX tiles of the southern Galactic disc, which includes 166 tiles from the disc-low,  
166 tiles from the disc-high, 
and 180 tiles from the disc+230 regions. Both surveys in these parts of the disc are deeper than the all-sky 2MASX survey and the 2MASS Redshift Survey \citep{Huchra2005,Huchra2012,Macri2019}, which did not reach this close to the Galactic Plane.  
Table 1 of \cite{Daza2023} provides a summary of both surveys and the different regions of the Galactic bulge and disc that were observed.

\begin{table*}
\caption{The Galactic disc regions covered by the VVV and VVVX surveys. 
 }
\begin{tabular}{llrccll}
\hline
\hline
Survey  & Disc       & Tiles &  Galactic  &  Galactic & Catalogues & Publication\\
        & region     &       &  longitude &  latitude &  &   \\
\hline
\hline           
VVV    &  \textit{disc}      & 152  &  295$^{\circ}$ $<$ $l$ $<$ 350$^{\circ}$   &  -2.25$^{\circ}$ $<$ $b$ $<$  +2.25$^{\circ}$ & VVV NIRGC I & \cite{Baravalle2021}\\
VVVX   &  \textit{disc+20}    &  56  &  10$^{\circ}$ $<l<$ 20$^{\circ}$           &  -4.5$^{\circ}$  $<b<$ +4.5$^{\circ}$   & VVV NIRGC II & \cite{Daza2023}\\
VVVX   &  \textit{disc-low}   & 166  &  230$^{\circ}$ $<l<$ 350$^{\circ}$         &  -4.5$^{\circ}$  $<b<$ -2.25$^{\circ}$  & VVV NIRGC III & This work\\
VVVX   & \textit{disc-high}   & 166  &  230$^{\circ}$ $<l<$ 350$^{\circ}$         &  +2.25$^{\circ}$ $<b<$ +4.5$^{\circ}$   & VVV NIRGC III & This work\\
VVVX   & \textit{disc+230}   & 180  &  230$^{\circ}$ $<l<$ 295$^{\circ}$         &  -2.25$^{\circ}$ $<b<$ +2.25$^{\circ}$  & VVV NIRGC III & This work \\
\hline
\label{tab:table0}
\end{tabular}
\end{table*}


\section{The galaxy detection procedure}\label{procedure}

In this study, we have employed a methodology developed by \cite{Baravalle2018,Baravalle2021} to identify galaxies behind the southern Galactic disc. 
We utilised $J$, $H$ and $K_{s}$ images of the 512 tiles within the VVVX survey covering this area. These images are available in the European Southern Observatory (ESO) archive\footnote{\url{https://doi.eso.org/10.18727/archive/68}}.
We selected images in the three passbands with exposures of
10, 6 and 4 seconds, respectively, and with the best available seeing in each passband. The median value of the seeing was 0.80~arcsec in the $K_{s}$ passband.

\subsection{Photometry and classification of extended sources}

We ran \dos on the selected images of $J$, $H$ and $K_{s}$ passbands across the 512 tiles, resulting in good determinations of the Point Spread Function (PSF) of the stars in each tile. The $K_{s}$ image was used as the reference, and all detected sources were cross-matched with the other two passbands using an angular separation of 1 arcsec. The \dos output is a catalogue with astrometric, morphological and photometrical parameters of the detected sources.  The morphological parameters were used to characterise the detections as point or extended sources. The \sex~configuration input, used in this paper, was slightly modified, mainly the parameters associated with the minimum number of pixels detected above the threshold, the minimum contrast for deblending objects and the background estimation. These changes were introduced taking into consideration that both interstellar extinction and stellar crowding are less severe in this part of the southern disc than in the inner region studied in \cite{Baravalle2021}.

The detected objects were divided into point and extended sources using a combination of four morphological parameters provided by {\sc SExtractor+PSFEx}: the CLASS\_STAR index, the SPREAD\_MODEL ($\Phi$) parameter, the half-light radius (R$_{1/2}$) and the concentration 
index (C, \citealt{Conselice2000}) defined as the ratio of two circular radii that contain 80\% and 20\% of the total Petrosian flux.
Furthermore, we performed a cross-match with Gaia-DR3 \citep[][]{2021A&A...650C...3G} considering an angular separation of $1.3$ arcsec between sources to remove additional point sources.
We obtained a total of $3,541,627$ extended objects that satisfied the following criteria: $\Phi > $ 0.002; 2.1 $< $  C $ < $ 5 \citep{Baravalle2018}; and,  depending on the $K_{s}$ magnitudes, for bright sources, $K_{s}$ $<$ 11 mag, R$_{1/2}$  $>$ 1 arcsec and CLASS\_STAR $<$ 0.1; for 11 $< K_{s}$ $<$ 14 mag, R$_{1/2} >$ 0.6 arcsec and  CLASS\_STAR $<$ 0.1; and for faint sources, 14 $ < K_{s}$ $<$ 17 mag, R$_{1/2} > $ 0.6 arcsec and CLASS\_STAR $<$ 0.3.

\subsection{Characterization of extended source classification}

In order to evaluate our ability to detect extended sources in the VVVX images, we calculated the completeness, or detection fraction,
by considering specific characteristics of the tiles, including the distribution and density of objects, as well as interstellar extinction. 
We used six tiles at different places on the Galactic disc with different interstellar extinctions. Per tile, 5,000 galaxies were injected, distributed randomly and homogeneously across the entire image area.
For this purpose, we generated synthetic galaxies modelled using the \texttt{Sersic2D} module from the \texttt{astropy.modeling.models}\footnote{\url{https://docs.astropy.org/en/latest/modeling/models.html}} package.
Each synthetic galaxy was characterized by different parameters that define both bulge and disc components. We included the bulge-to-total luminosity ratio, ranging between 0 for pure disc galaxies to 1 for pure bulge galaxies. We also included the bulge-specific parameters as effective radius, ranging from 1 to 5 pixels; aspect ratio derived from the average ellipticity measured in the image and position angle defined within the range $0$ to $\pi$ radians and the disc-specific parameters as scale length,  varying from 5 to 20 pixels; inclination angle from $0$  to $\pi/2$ for face-on to edge-on galaxies, respectively; and the disc position angle defined between $0$ and $\pi$ radians. Magnitudes followed a Beta-type distribution that spanned magnitudes of 8 to 22 mag, with parameters $\alpha = 8$ and $\beta = 2$ defining both the peak and the tail characteristics of the distribution.
The magnitudes were derived using the photometric zero-point provided in the image header and taking into account the mean interstellar extinction estimated for the tile.
Each galaxy was subsequently convolved with a Gaussian kernel, the size of which matches the characteristic Full Width at Half Maximum of the observational image. This convolution ensured that the synthetic galaxies possess photometric and structural properties consistent with observed galaxies, thus enabling a robust evaluation of galaxy detections.

We run \dos over the synthetic $K_{s}$ images of the tiles 
and the generated catalogue was used to obtain the percentage of detections of synthetic extended sources. 
Figure~\ref{fig:completeness} shows the completeness results for two tiles (e1114 and e0605) with median interstellar extinctions A$_{K_s}$ of 0.78 mag and 0.32 mag, respectively. The results for the other tiles are included in the grey shadow region.
We find a completeness of 80\% between 15.5 to 16.5 mag in the $K_s$ magnitudes.
The observed differences in the completeness curves are related to both interstellar extinctions and the density of the sources. 
These results are in agreement with those of the VVV survey:  \cite{Saito2012} for  CASU photometry and \cite{Baravalle2018} for \dos photometry for star and galaxy detections.

\begin{figure}[ht]
\centering
\includegraphics[width=0.4\textwidth]{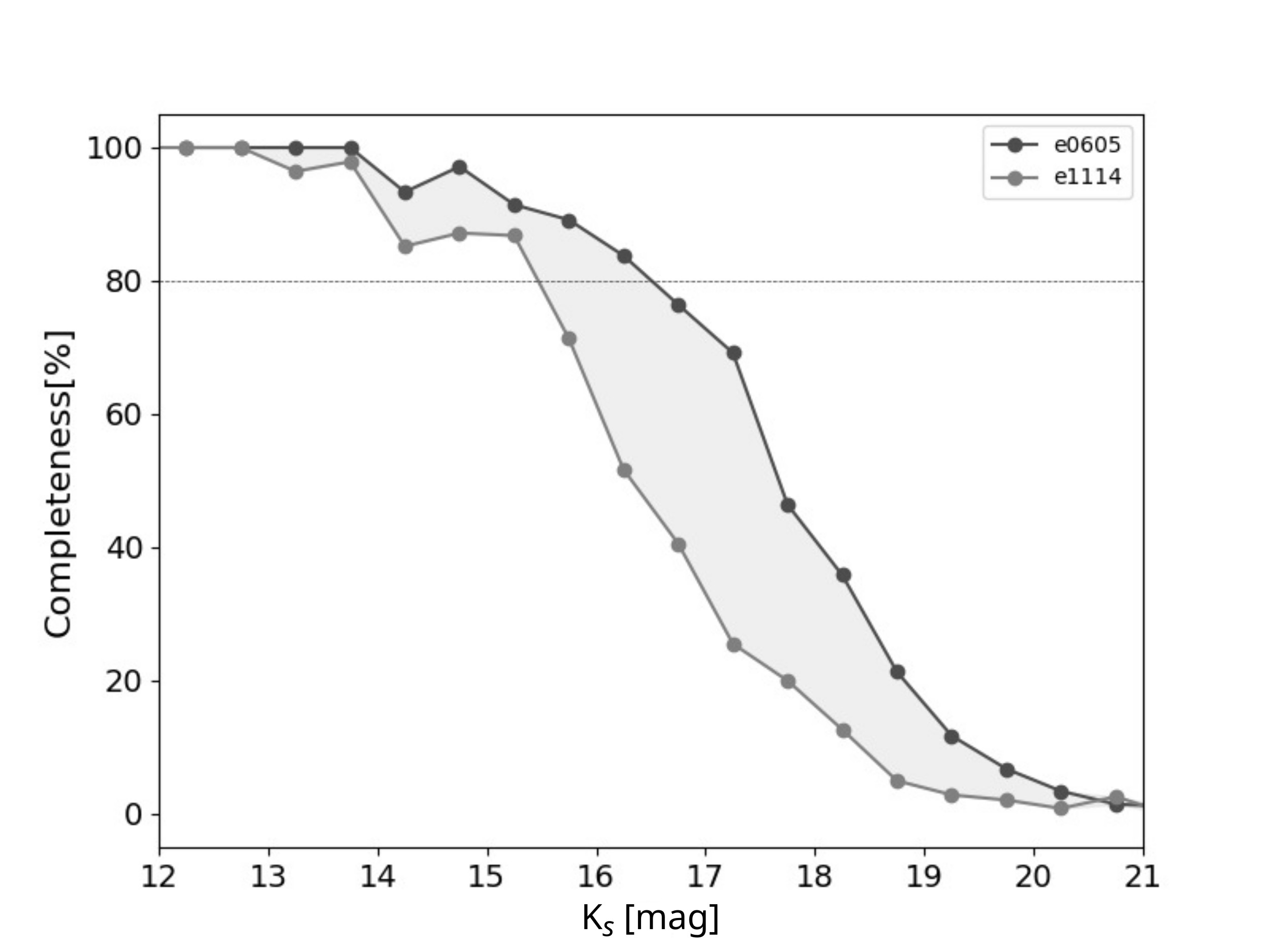}
\caption{Completeness in percentage for input $K_{s}$ magnitudes of synthetic galaxy detections. The grey and black dots, along with the solid line, correspond to the results of the e1114 and e0605 tiles, respectively. The 80\%
completeness level is also shown.}
\label{fig:completeness}
\end{figure}

We limited our characterization of detection efficiencies to the procedure described above, and did not extend it to the subsequent classification detailed below in \S3.3. A full evaluation of that step in our analysis would require the use of a more sensitive, reliable, and independent data set \citep{2000AJ....119.2498J}. Our study is the first of its kind at these low Galactic latitudes in the ZoA disc, and this region exhibits rapid changes in extinction and stellar density. Simulating these effects on a population of synthetic galaxies to determine their impact on observed colours and the parameters used by our adopted machine-learning techniques lies beyond the scope of this work.

\subsection{Classification of possible extragalactic sources}

To distinguish between Galactic and
possible extragalactic sources, we applied the colour selection criteria as in \cite{Baravalle2021}, namely: 0.6 $<$ ($J$ - $K_{s}$)$^{\circ}$ $<$ 2.0 mag,
0.0 $<$ ($J$ - $H$)$^{\circ}$ $<$ 1.0 mag,  
0.0 $<$ ($H$ - $K_{s}$)$^{\circ}$ $<$ 2.0 mag, and ($J$ - $H$)$^{\circ}$ + 0.9 ($H$ - $K_{s}$)$^{\circ}$ $>$ 0.44 mag. These colour cuts were defined using the colour-magnitude and colour-colour diagrams where the different extended objects could be identified. We also supported the defined colour limits with visual inspection of the objects trying to minimize the exclusion of genuine galaxies. We obtained a total of $723,637$ sources. 
We also discarded objects found at the edges of the tiles and duplicate sources in general, leading to a final number of $692,694$  possible extragalactic sources. 
Among these sources, we recognize the presence of both genuine galaxies and false detections. While this task was previously performed by visual inspection \citep{Baravalle2018,Baravalle2021}, we decided to employ machine learning techniques as done in \cite{Daza2023} to get rid of false detections.  
Given the excellent results obtained in the analysis of the northern Galactic disc and the increase in the number of tiles in VVVX coverage, we used these algorithms in order to identify galaxies behind the southern Galactic disc.

Two approaches were used: the first utilised image information from the
$J$, $H$ and $K_s$ passbands, creating the Image Sample (IS) with stamps of 44 $\times$ 44 pixel spatial size for supervised methods. The second approach, the Photometric Sample (PS), used photometric and morphological information from the \dos outputs.  
The data classification procedure in this work is analogous to that used on the northern Galactic disc \citep{Daza2023} in terms of data curation and pre-processing, feature generation and selection in both approaches. In addition to the labeled samples from the northern disc, a labeled sample of galaxies (Gx) and non-galaxies (non-Gx) was generated by visual inspection of $23$ tiles from the VVVX coverage of the southern Galactic disc to perform transfer learning on data from these new regions. These tiles contain interesting objects such as groups and clusters of galaxies in different parts of the sky. These objects were divided into a training set and a test set, containing $70$\% and $30$\% of the sample, respectively. In this way, the southern disc training set consisted of $24,150$ sources, of which $12,771$ were labeled as non-Gx and $11,379$ as Gx, while the test set contained a total of $10,350$ sources composed of $5,474$ non-Gx and $4,876$ Gx. 
In the case of IS, the probability that an object is a galaxy was determined using CNN with $44 \times 44 \times 6$ images as input to the model. Three of the six images corresponded to the sources in the $J$, $H$, and $K_s$ passbands, individually scaled, while the remaining images were generated by applying an edge detection and smoothing filter, known as Gauss + Laplace, to the images in each passband.
In the PS case, the probabilities were calculated using XGBoost, which was trained with 
$58$ photometric and morphological features. After applying the ML algorithms, the CNN based on the images (IS-CNN) and XGBoost trained with the photometric and morphological data (PS-XGBoost), the probabilities were obtained for all possible extragalactic sources.

The performance of these models is typically evaluated using precision and recall metrics.
The precision metric is a measure of the ability of the ML models to correctly classify galaxies. It is defined as $\frac{\text{TP}}{\text{TP + FP}} $, where TP (true positive) is the number of objects correctly classified as Gxs by the ML algorithms and FP (false positive) is the number of objects representing non-Gx, incorrectly classified as Gxs.  The recall metric is the fraction of galaxies identified by the models and thus it gives us an insight into the completeness of the classification. It is defined as
$\frac{\text{TP}}{\text{TP + FN}}$, where FN (false negative) sources are Gxs classified by the ML algorithms as non-Gxs. A better precision or recall can be achieved depending on the chosen probability threshold. An in-depth analysis of the compromise between the two metrics can be done by examining the recall-precision curves, which show the values of these metrics for different thresholds calculated using the test set. Figure \ref{fig:recall_precision} shows these curves for our two used ML models, IS-CNN and PS-XGBoost, colour coded for different threshold values, with the most commonly used value of 0.5 marked with red crosses as a reference.

\begin{figure}[ht]
\centering
\includegraphics[width=.46\textwidth]{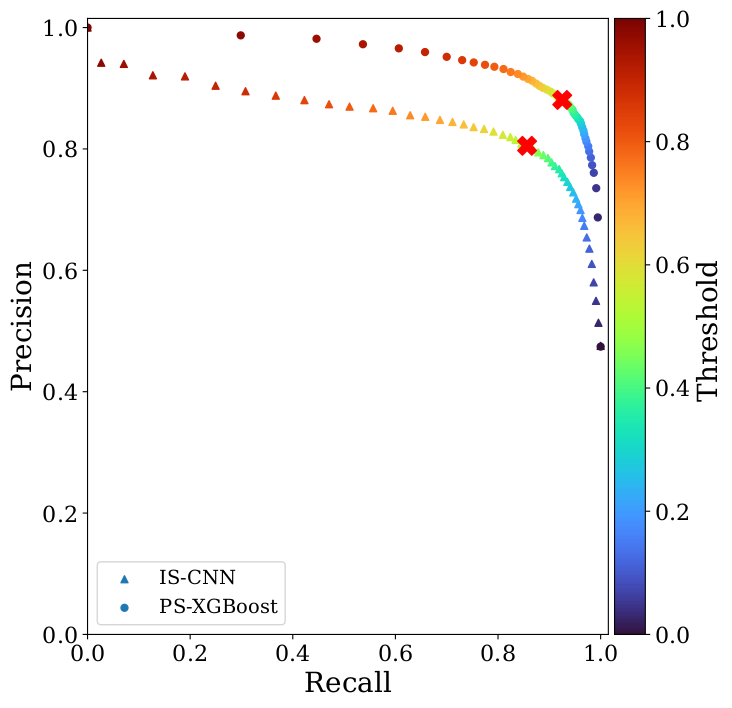}
\caption{Recall-Precision curves for the IS-CNN and PS-XGBoost models. The red crosses indicate the most common threshold value of 0.5. Other threshold values are indicated by the colour bar.}
\label{fig:recall_precision}
\end{figure}

A classifier that exhibits both high precision and low recall will return a low number of objects, yet almost all of these will be correctly labelled. Conversely, a system that shows low precision and high recall will recover the majority of true positive and negative sources; however, the proportion of incorrectly labelled results will be high. For this work, we prioritize a better precision than recall in order to emphasize purity over completeness.


\section{The VVV near-IR galaxy catalogue III\label{sec:catalogue}}

We applied the improved procedure and obtained 692,694 possible extragalactic sources with their
probabilities of being galaxies using machine learning techniques.  
Of the duplicates found in the overlap regions of both VVV and VVVX surveys, we
only kept the measurements from the VVV survey.
The magnitudes and colours of the sources were corrected for interstellar extinction along the line of sight using the maps of \cite{Schlafly2011} obtained via the dustmaps\footnote{\url{https://dustmaps.readthedocs.io}} library for \texttt{Python} \citep{dustmaps}, with an interpolation of order 2.  
The VVV NIR relative extinction coefficients used are those of \cite{Catelan2011}. 

It is not easy to accurately calculate the total magnitude of galaxies at low latitudes. We used the automatic magnitude (MAG\_AUTO) computed using the Kron method \citep{Kron1980}, which defines an elliptical aperture that fits the shape of the source and adapts
to its size and orientation. On the other hand, isophotal magnitudes are computed in an aperture defined by an isophotal contour determined by a fixed intensity threshold. This threshold is obtained as a multiple of the background standard deviation, which could be affected by the flux of adjacent sources, especially if they are close, making MAG\_AUTO 
 more reliable under these conditions.
All magnitudes were transformed to the 2MASS photometric system following the same procedure as described by \cite{Soto2013}. Briefly, photometric transformations were calculated on a tile-by-tile basis as follows: From the $JHK_s$ catalogue of aperture photometry, stars in close proximity to the brightest star and with a magnitude difference in $K_s$  of less than 2 magnitudes were discarded. The remaining stars were cross-matched with the 2MASS point source catalogue \citep{Cutri2003} in the same tile area, retaining only those with quality flags ‘A’ or ‘B’ within a radius of 0.3 arcsec. Finally, linear fits were calculated using the resultant cross-referenced VVV-2MASS catalogue, employing an iterative clipping algorithm to reject outliers beyond 2.5 $\sigma$.  

The near-IR galaxy catalogue from our analysis of VVVX coverage of the southern Galactic disc, VVV NIRGC III, contains 167,559 galaxies defined with the IS-CNN and PS-XGBoost joint probabilities higher than 0.6, or with IS-CNN probabilities between 0.5 and 0.6 and PS-XGBoost probabilities higher than 0.8. These probability limits were determined through the analysis of the recall-precision curves in Figure \ref{fig:recall_precision} and a meticulous visual inspection of several sources, and they are regarded as conservative. The compromise between contamination and completeness is not a simple one, but we defined our selection criteria prioritizing purity. These joint probability limits yield a precision of 0.91 and a recall of 0.78 for the test set. These values indicate a contamination rate due to misclassifications of less than 10\% in our final catalogue and a completeness of 78\%.
From the catalogue, 16,998 galaxies were visually confirmed in regions characterized by relatively mild interstellar extinctions, in contrast to the other two previously studied regions of the Galactic disc. 

Based on the visual inspection of sources made post training and classification, excluding the training and test sets, we also estimated the recall and precision metrics. We obtained a 12\% contamination rate and 80\% completeness, which are in good agreement with metrics from the test set. For large sources (R$_{1/2} \ge $ 3.5 arcsec), we found a completeness of 99.7\% and a contamination of 17\% mostly corresponding to Galactic objects and image artifacts like spikes. Figure~\ref{fig:false} shows VVVX colour composed images of examples of these false detections. We have
333 large galaxies visually checked in our final catalogue. Sources with smaller R$_{1/2}$ (0.8 $<$ R$_{1/2} <$ 1.0 arcsec) show a completeness of about 83\% with similar contamination rates. For even smaller objects (0.6 $<$ R$_{1/2} <$ 0.8 arcsec), the contamination rates remain similar but the completeness drops to 61\% mainly because neither VVV NIRGC I nor VVV NIRGC II included objects with R$_{1/2} <$ 1.0 arcsec and thus transfer learning is less efficient in these cases. Furthermore, these are fainter and more diffuse objects, leading to a harder classification. Even with these concerns, we decided to include these smaller objects in our final catalogue given that visual inspection has confirmed that these sources are indeed galaxies, including 8,991 confirmed galaxies in this half-light radius range.

\begin{figure*}[ht]
\centering
\includegraphics[width=0.26\textwidth]{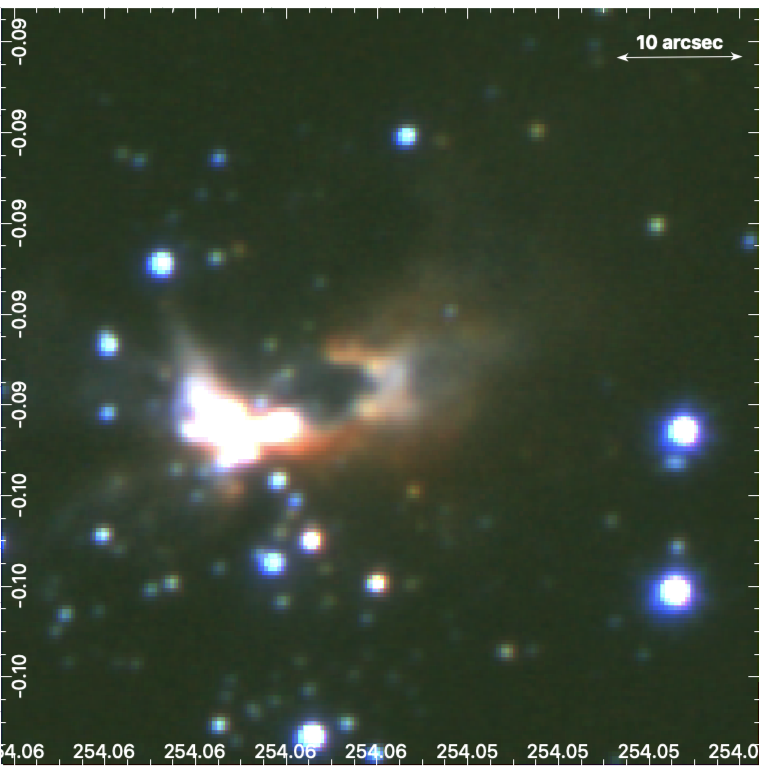}
\includegraphics[width=0.26\textwidth]{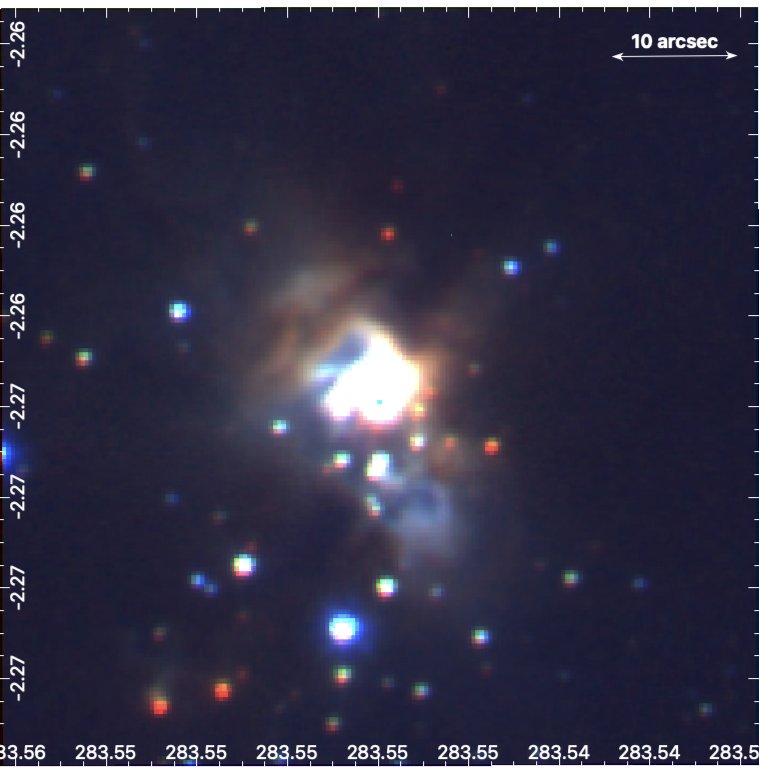}
\includegraphics[width=0.26\textwidth]{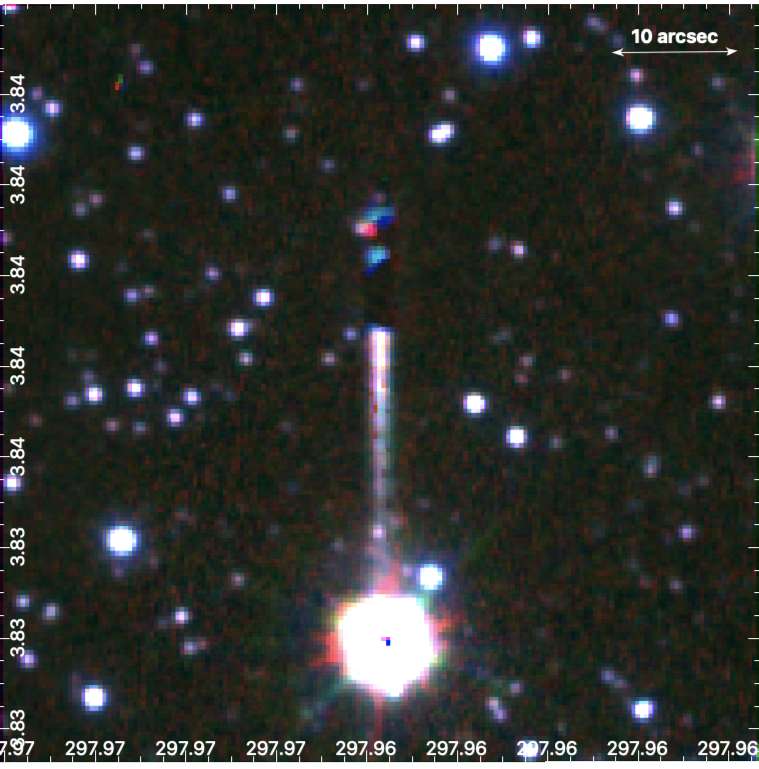}
\caption{Examples of false detections. From left to right panels, two Galactic star forming regions and a spike are shown in the central parts of the VVVX images. North is up and east is to the left.} 
\label{fig:false}
\end{figure*}

Table~\ref{tab:catalogue} shows the properties of the first ten sources of the catalogue including the identification in column (1), the J2000 equatorial coordinates in columns (2) and (3), the Galactic coordinates in columns (4) and (5), the A$_{Ks}$ interstellar extinction in column (6), total extinction-corrected $J^{0}$, $H^{0}$, and $K_{s}^{0}$ magnitudes in columns (7) to (9), the extinction-corrected $J_{2}^{0}$, $H_{2}^{0}$, and $K^0_{s2}$ aperture magnitudes within a fixed aperture of 2~arcsec diameter in columns (10) to (12), and the morphological parameters $R_{1/2}$, $C$, ellipticity and $n$ in columns (13) to (16), respectively where  $R_{1/2}$ is the half-light radius in arcsec, $C$ is the concentration index, 
 and $n$ is the spheroid Sersic index \citep{Sersic1968}. The probabilities of the IS-CNN and PS-XGBoost models are shown in columns (17) and (18), respectively, and the galaxy and redshift flags in columns (19) and (20), respectively. In the galaxy flag column, a value of 1 indicates that the source has undergone a visual inspection, including sources used to train the ML algorithms and those inspected at a later stage, and 2 for 2MASX galaxies. In the training cases, 
the IS-CNN and PS-XGBoost probabilities are set to 1 for galaxies (0 for non-galaxies). For the redshift flag column, 1 and 2 indicate galaxies with available spectroscopic and photometric redshifts, respectively. This catalogue is available in electronic format \footnote{\href{https://catalogs.oac.uncor.edu/vvvx_nirgc/}{VVV\_NIRGC\_III.csv}}. 
The complete catalogue of 692,694 possible extragalactic sources, along with their probabilities of being galaxies is available upon request. Also, aperture magnitudes within aperture diameters of 4, 6 and 7 arcsec are available upon request.   

Figure~\ref{fig:histoAKs} shows the distribution of the interstellar extinctions in the $K_{s}$ passband for the VVV NIRGC catalogues,
in bin counts normalized with the total sample size (N). It is evident from this figure that the regions of VVV NIRGC I, or the inner parts of the southern Galactic disc, are the most extincted regions. These start with values higher than 0.2 and reach 3.0 magnitudes. In contrast, the southern disc regions covered by VVV NIRGC III have the smallest values and peaks of around 0.25 magnitudes. The northern disc regions of VVV NIRGC II have intermediate extinction values between the other two distributions.

\begin{figure}[ht]
\centering
\includegraphics[width=0.9\columnwidth]{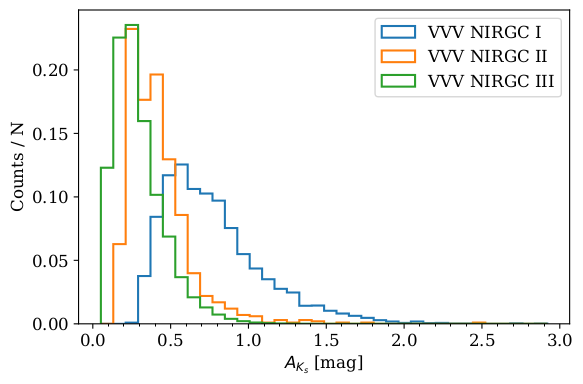}
\caption{Distributions of interstellar extinctions in the $K_{s}$ passband for the VVV NIRGC catalogues. Bin counts are normalized with the total sample size. Blue, orange and green lines are those to denote the distributions related to the catalogues VVV NIRGC I, II and III, respectively. }
\label{fig:histoAKs}
\end{figure}

\begin{table*}
\small
\caption{The VVV NIRGC III galaxy catalogue from the VVVX coverage of the southern Galactic disc with their probabilities. The full table is available at the CDS.}
\begin{tabular}{lccccccccccccccccccc}
\cline{1-9}
\noalign{\vskip\doublerulesep
         \vskip-\arrayrulewidth}
\cline{1-9}
ID & RA &  Dec & l & b & A$_{Ks}$ & $J^0$  &  $H^0$ & $K_s^0$ \\
 & (J2000) &  (J2000) & [$^{\circ}]$ & [$^{\circ}$] &  [mag]  &  [mag] &  [mag]  &  [mag]  \\
\cline{1-9}
\noalign{\vskip\doublerulesep
         \vskip-\arrayrulewidth}
\cline{1-9}
VVVX-J071609.20-195027.7  &  07:16:09.20  & -19:50:27.7  & 233.356   &   -3.707  &  0.34     &  11.43      &   10.75     &   10.67  \\
VVVX-J071848.37-191254.5  & 07:18:48.37   & -19:12:54.5  &  233.092  &  -2.863   &  0.38     &  12.70       &   12.11     &   12.22  \\
VVVX-J072520.07-245648.2  & 07:25:20.07   & -24:56:48.2  &  238.875  &  -4.210   &   0.34    & 12.51       &   11.95     &   12.03 \\ 
VVVX-J072545.35-175253.0  & 07:25:45.35   & -17:52:53.0  &  232.689  &  -0.782   &  0.44     & 10.91       &   10.24     &   10.33   \\
VVVX-J072602.10-171749.7  & 07:26:02.10   & -17:17:49.7  &  232.208  &  -0.446   &  0.57     & 12.38       &   11.69     &  11.55  \\  
VVVX-J072646.30-235109.9  & 07:26:46.30   & -23:51:09.9  &  238.063  &  -3.405   &  0.38     & 11.16       &   10.50      &   10.53  \\
VVVX-J072729.21-235733.6  & 07:27:29.21   & -23:57:33.6  &  238.235  &  -3.312   &   0.35    & 10.84       &   10.15     &  10.09  \\
VVVX-J072949.71-253457.3  & 07:29:49.71   & -25:34:57.3  &  239.919  &  -3.615   &   0.29    &  11.64      &    10.94    &   10.97 \\  
VVVX-J073026.88-133828.5   & 07:30:26.88  & -13:38:28.5  & 229.501   &  2.233    &   0.12    & 12.82       &  12.11      &   12.07  \\  
VVVX-J073131.55-134811.8   & 07:31:31.55  & -13:48:11.8  & 229.769   &  2.385    &   0.14    &  13.56      &  12.91      &   12.88  \\  
                    &              &              &
&           &           &             &             &
                      \\
\cline{2-12}
\noalign{\vskip\doublerulesep
         \vskip-\arrayrulewidth}
\cline{2-12}
  &  $J_2^0$ &    $H_2^0$ &  $K^0_s{}_2$ & $R_{1/2}$ & C &  $e$  & $n$ &  IS &  PS & Gx & z\\
  &   [mag] &   [mag] &   [mag]    &  [arcsec]  &   &       &     &     &  & flag & flag \\
\cline{2-12}
\noalign{\vskip\doublerulesep
         \vskip-\arrayrulewidth}
\cline{2-12}
&  14.08 &     13.32   &     13.03      &     4.05    & 3.48  & 0.51    &      4.31  & 1.0     &  1.0    &   1 & 2 \\    
&  15.33 &     14.54   &     14.41      &     3.82    & 3.11  & 0.41    &      5.33  & 1.0     &  1.0    &   1 & 2 \\
&  15.91 &     15.07   &     14.91      &     3.85    & 2.57  & 0.50    &      1.93  & 1.0     &  1.0    &   1 & 0 \\ 
&  14.31 &     13.49   &     13.25      &     6.18    & 3.16  & 0.32    &      4.53  & 1.0     &  1.0    &   1 & 1 \\
&  15.00 &     14.21   &     13.91      &     3.85    & 3.30  & 0.78    &      1.77  & 1.0     &  1.0    &   1 & 0\\ 
&  13.84 &    12.94    &     12.80      &     3.82    & 3.40  & 0.20    &      3.74  & 1.0     &  1.0    &   1 & 2 \\ 
&  14.36 &    13.49    &     13.29      &    6.81     & 3.60  & 0.64    &      2.54 &  1.0     &  1.0    &   1 & 1\\ 
&  14.31 &    13.37    &     13.22      &    4.02     & 3.32  &  0.05   &      5.31 &  1.0     &  1.0    &   1 & 2 \\
&  15.26 &    14.34    &     14.14      &    3.52     & 3.53  & 0.71    &      3.67 &  1.0     &  1.0    &   1 & 2\\  
&  16.66 &    15.81    &     15.65      &    3.60     & 2.60  & 0.69    &      0.97  & 1.0     &  1.0    &   1 & 1\\   
\cline{2-12}
\end{tabular}
\label{tab:catalogue}
\end{table*}

\subsection{Estimating errors}

The VVV and VVVX surveys exhibit tile overlaps. 
In these overlap regions, while image quality may be a concern, leveraging the detected sources for magnitude comparison would be advantageous, under the assumption that the intrinsic errors are smaller at greater distances from the image edges. 
Also, depending on 
size, it can be difficult to reach the fainter outer parts of the objects, thereby making comparison challenging. 
Figure~\ref{fig:diffVVVX_X} shows the differences in the total magnitudes for the three passbands as a function of the adopted magnitudes for the 20,438 pairs of extragalactic candidates observed in different tiles. The medians and standard deviations for the comparisons are $\Delta\, J   =  0.000 \pm 0.210$, 
$\Delta\, H   = -0.004 \pm 0.184$ and 
$\Delta\, K_s = -0.002 \pm 0.207$ mag. These differences reveal intrinsic dispersions because the sources were obtained using exactly the same procedure. Also, there were 441 galaxies in common between the VVVX and VVV surveys. The differences between the VVVX and VVV magnitudes have medians of $\Delta\, J   = -0.610 \pm 0.458$, $\Delta\, H   = -0.530 \pm 0.334$ and $\Delta\, K_s = -0.250 \pm 0.330$ mag.    
There was a smaller number of galaxies in common, and these higher median values and dispersions might be related to differences in the observations and the size of the galaxies. The most significant differences were found for larger galaxies always observed at the tile boundaries.

\begin{figure*}[ht]
\centering
\includegraphics[width=1\textwidth]{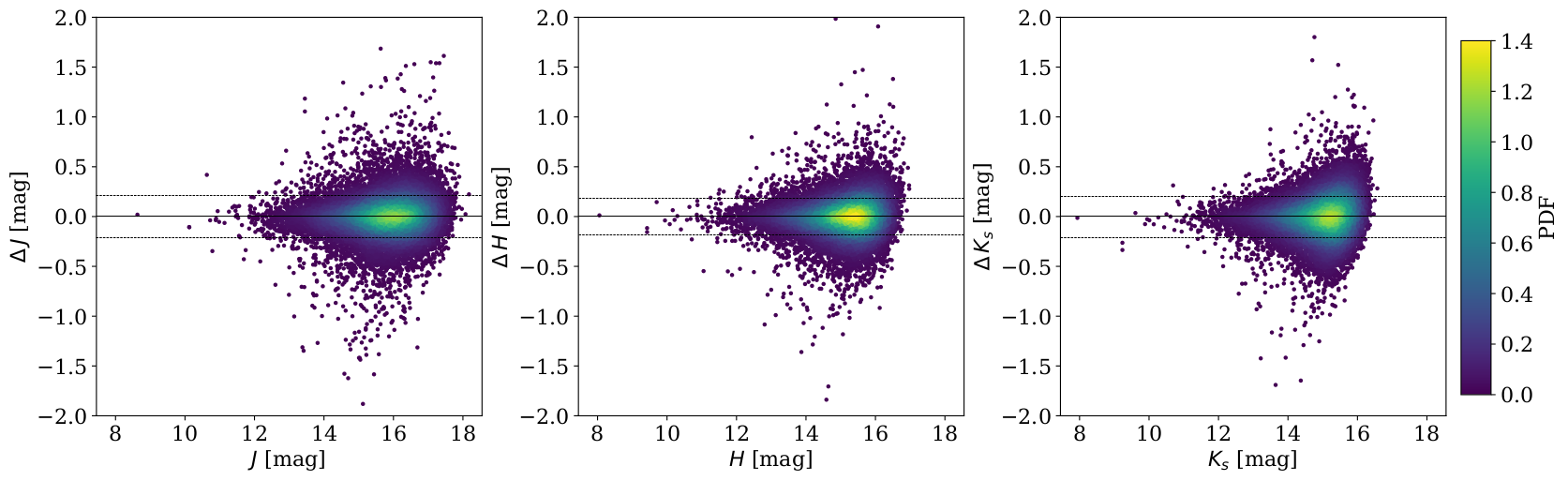}
\caption{Internal magnitude comparisons in the three NIR passbands of the VVVX survey. The colour bar indicates the probability density function (PDF) estimated using a Gaussian kernel density estimation. } 
\label{fig:diffVVVX_X}
\end{figure*}

\subsection{Multi-wavelength counterparts}

\subsubsection{Photometric surveys}

The VVV NIRGC III was cross-matched with 2MASX \citep{Jarrett2004}, revealing a total of 6,058 sources in common. 
As mentioned above, all of our NIR magnitudes were transformed to the 2MASS photometric system. This permits a magnitude comparison, thus ensuring the reliability of our results. Figure~\ref{fig:VVVX_2MASX} shows the comparison $\Delta$ mag = mag(VVVX) - mag(2MASX) between total magnitudes in the three NIR passbands.  
The medians and standard deviations are: 
$\Delta\, J   = -0.72 \pm 0.51$, $\Delta\, H   = -0.45 \pm 0.38$ and
$\Delta\, K_s = -0.02 \pm 0.34$.  
The limiting magnitudes for extended objects in the 2MASX are 15.0, 14.3, and 13.5 mag for the $J$, $H$ and $K_s$ passbands, respectively. Consequently, the distributions are asymmetric for fainter magnitudes where selection effects and uncertainties are more significant. 
Also, we compared our total magnitudes with 
the deep NIR photometry of \cite{Williams2014} with the same photometric procedure as 2MASX without transforming magnitudes to the 2MASS system. The median of the magnitude difference in the $K_s$ passband is 
$\Delta\, K_s = 0.26 \pm 0.70$ mag using 128 galaxies in common.  
 The comparisons are more robust than in \cite{Baravalle2021} with an important number of objects in common and the differences observed  can be attributed to the different photometric methods and procedures used to define {total magnitudes.

In the mid-IR passbands, VVV NIRGC III was cross-matched with 
the Wide-field Infrared Survey Explorer mission (WISE, \citealt{Wright2010}), producing 90,123 galaxies in common. 
The WISE survey detects a wide range of galaxies in the Universe. However, its sensitivity to mid-IR emission makes it more efficient in identifying luminous galaxies, often associated with star formation, active galactic nuclei (AGN) activity \citep{Massaro2012}, or mergers \citep{Shanshan2013}.
A multi-frequency analysis using data from high-energy sources combining near- and mid-infrared surveys, allowed us the identification of AGN behind the MW at low latitudes ($ -10^\circ < b <$ 3$^\circ$) \citep{Pichel2020,Baravalle2023,Donoso2024}.

\begin{figure*}[ht]
\centering
\includegraphics[width=1\textwidth]{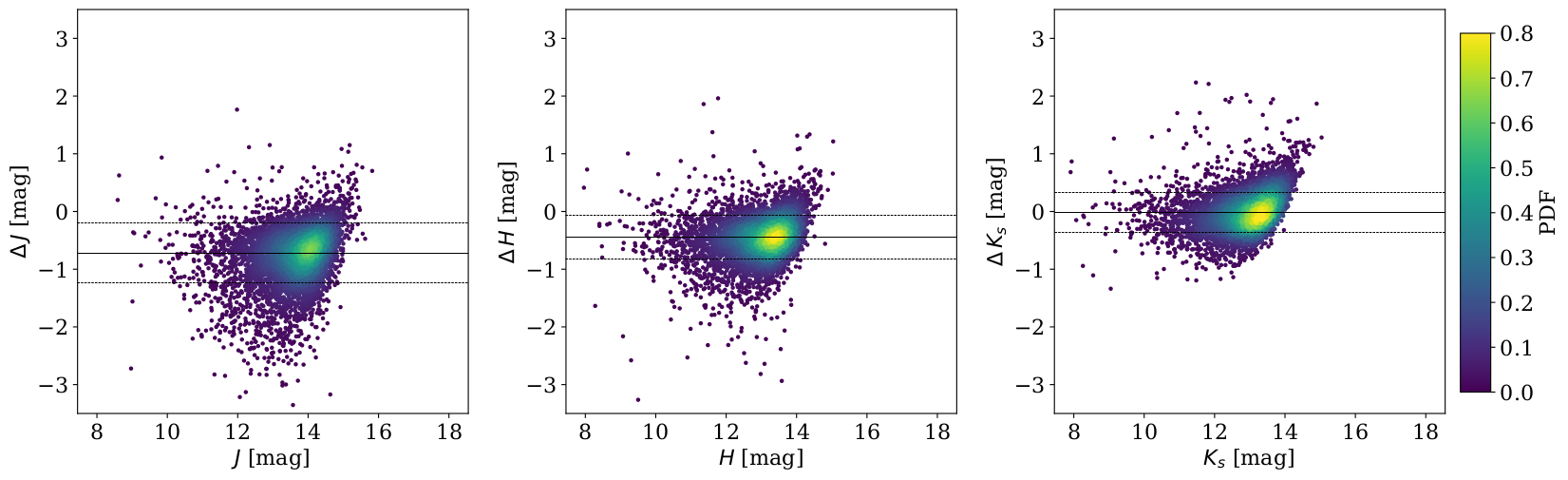}
\caption{The comparison between the total magnitudes of VVV NIRGC III  and the 2MASX in the three NIR passbands. The colour bar indicates the PDF estimated using a Gaussian kernel density estimation, as shown in Fig.\ref{fig:diffVVVX_X}}
\label{fig:VVVX_2MASX}
\end{figure*}

\subsubsection{Galaxies with redshift information}
  
The VVV NIRGC III catalogue was cross-matched with the Southern Hemisphere Spectroscopic Redshift Compilation \citep{delima_specz_compilation}, which contains more than 1,700 spectroscopic redshift catalogues, in order to include as many galaxies as possible with available redshift data. Using  an angular separation  of 2 arcsec, this cross-match resulted in a total of 286 galaxies from the Local AGN survey (LASr, \citealt{Asmus2020}); the Galaxy groups catalogue within 3500 km/s \citep{Kourkchi2017}; Parkes HI observations of galaxies behind the southern MW (Crux and Great Attractor regions, \citealt{Schroder2009});  structures in the Great Attractor region \citep{Radburn2006}; the Deep NIR photometry of HI galaxies in the ZoA \citep{Williams2014}; redshift distribution of galaxies in the Southern MW region \citep{Visvanathan1996} and  GAIA DR3 \citep{2020yCat.1350....0G} among the most important references.  
Furthermore, using the same angular separation of 2 arcsec with SIMBAD database there are 30 additional galaxies with spectroscopic radial velocities in the region mainly from the 2MASS bright galaxies ZoA survey in HI \citep{Kraan2018}.

The SARAO MeerKAT galactic plane survey \citep{Goedhart2024} provides HI images with an angular resolution of up to 30 arcsec. The VVV NIRGC III catalogue was cross-matched with the works of \cite{Rajohnson2024,2024MNRAS.535.3429R}} using an angular distance of 10 arcsec based on the positional accuracy of the sources. 
A total of 108 and 96 galaxies from these two studies, respectively, have been identified within the Vela Supercluster region.

Furthermore, VVV NIRGC III was cross-matched with the 2MASS photometric redshift catalogue \citep{Bilicki2014} using an angular separation of 2 arcsec. This catalogue includes photometric and spectroscopic radial velocities. We have in common 3,670 galaxies with photometric redshifts, with only 6 galaxies with both photometric and spectroscopic measurements.

The total number of redshifts available in the literature without duplicates for the southern Galactic disc is 493 for galaxies with spectroscopic redshifts and 3,670 for galaxies with photometric redshifts. These represent approximately 2.5\% of the VVV NIRCG III catalogue. The median values of the redshift distributions are $z=0.02$ $\pm$ $0.03$ and $z=0.06$ $\pm$ $0.03$ for spectroscopic and photometric redshifts, respectively, indicating that these galaxies belong to the local Universe. Both distributions reach values up to 0.2. 
It is imperative to obtain both photometric and spectroscopic redshifts to have an extensive sample of galaxies, including distant ones, to enhance our understanding of the galaxy distribution in this region of the ZoA, thereby contributing to the discernment of the role of the local Universe in the determination of the Hubble constant. Mapping of galaxies together with their distances can help to understand the dynamics of the local Universe in both denser regions and voids.

\subsubsection{Additional sources in the region}

There are 4,314 extended sources in 2MASX without cross-matches in
VVV NIRGC III. Most of these sources are Galactic objects, including nebulae and young stellar objects, which are located in regions of high star formation.
Additionally, there are 43 galaxies from the 2MASX catalogue in the regions of the southern Galactic disc with available radial velocities not included in VVV NIRGC III. The majority of these sources are located behind regions of active star formation. 
In 11 cases, the $J$ and/or $H$ VVVX images were insufficiently bright to yield reliable photometric parameters. Four galaxies were rejected due to their morphology and 14 more because of their colours. Also, there were 14 large galaxies that did not satisfy some of our criteria and so they were not included in the catalogue. Some examples of these are 2MASX J09023729-4813339 \citep{Zurita2009}, 2MASX J14130990-6520204 \citep{Freeman1977}, 2MASX J15523438-5823397 \citep{Visvanathan1996} and Circinus \citep{Freeman1977}.


\section{Analysis of the VVV NIRGC catalogues}
\label{sec:analysis}

Table~\ref{tab:summary} shows the summary of the VVV NIRGC catalogues in the three studied regions of the Galactic disc including  the median of the interstellar extinctions in the $K_{s}$ passband (related to the distributions shown in Figure~\ref{fig:histoAKs}),
the total number of sources detected by {{\sc SExtractor+PSFEx}}, 
and the extended and possible extragalactic sources after applying the morphological and colour criteria.
We also included the total number of galaxies in the VVV NIRGC catalogues and those in common with other NIR photometric surveys.

The total number of identified objects in VVV NIRGC I was higher despite the smaller number of tiles. This was primarily due to the fact that our study focused on the inner regions of the Galactic disc, where the density of stars and the crowding of the area were the most prominent characteristics. While the number of survey tiles in VVV NIRGC III was over three times that of VVV NIRGC I, the total number of identified objects was slightly lower, consistent with the lower density of stars in the disc. Conversely, the number of galaxies observed in VVV NIRGC III} was significantly increased. 

\begin{table*}
\caption{Statistics of the VVV NIRGC catalogues.}
\begin{tabular}{lllrrrrr}
\hline
\hline
Catalogue & Region & A$_{Ks}$ &  Total & Extended & Possible Extragalactic & Galaxies & Published \\
 & & & sources & sources & sources &  & NIR surveys \\
\hline
\hline
NIRGC I & VVV: southern disc & 0.71$\pm$0.33 & 177,838,607 & 2,070,768 & 80,038 & 5,554  & 45 \\ 
NIRGC II & VVVX: northern disc  & 0.48$\pm$0.21 & 66,983,004  &   871,071 & 172,396 &  1,003  & 2 \\ 
NIRGC III & VVVX: southern disc & 0.26$\pm$0.16 & 107,135,095 & 3,541,627 & 692,694 & 167,559 & 6,058 \\ 
\hline
\end{tabular}
\label{tab:summary}
\end{table*}

\begin{figure*}[ht]
\centering
\includegraphics[width=1\textwidth]{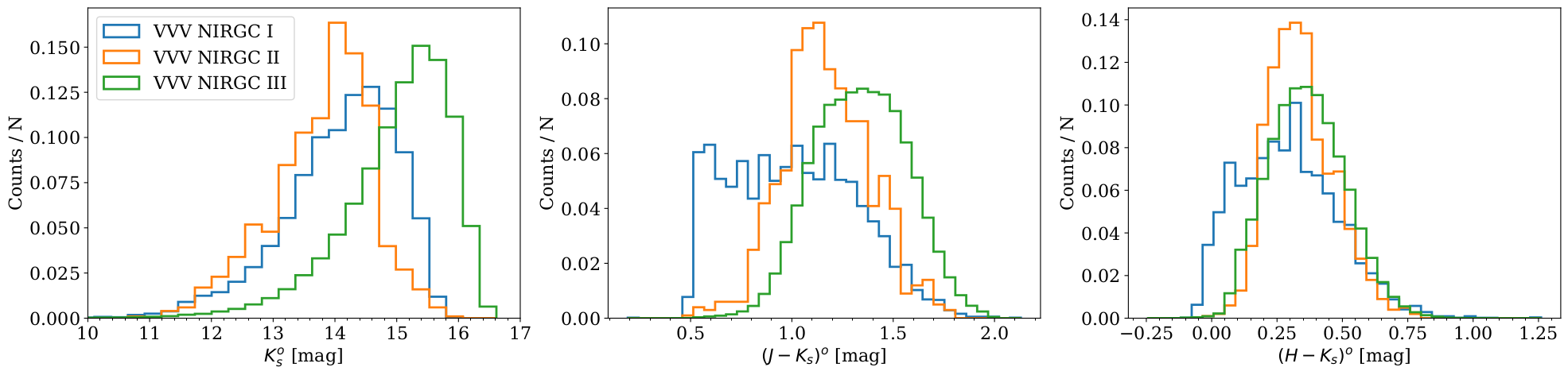}
\caption{Distributions of the photometric parameters  $K^0_{s}$ magnitudes and colours $(J - K_s)^0$ and $(H - K_s)^0$.}
\label{fig:histophot}
\end{figure*}

Figure~\ref{fig:histophot} shows the distributions of photometric parameters such as extinction-corrected $K^0_{s}$ magnitudes and colours $(J - K_s)^0$ and $(H - K_s)^0$. For VVV NIRGC I and II, the $K^0_{s}$ distributions are similar, showing a peak at about 14 mag and reaching up to 16 mag, while VVV NIRGC III shows a peak at about 16 mag and reaches up to 17 mag. This result is in agreement with the completeness of 80\% in $K^0_{s}$ magnitudes in the range of 15.5 to 16.5 mag (Figure~\ref{fig:completeness}). 
The $(J - K_s)^0$ colour distribution of VVV NIRGC I is distinct from those of the other two regions, exhibiting broader distribution. In these regions of high interstellar extinction and stellar crowding, galaxies are fainter with higher uncertainties and, for bluer colours there is about 30\% of contamination,  primarily attributed to the presence of nearby stellar light. The VVV NIRGC III catalogue also contains a higher number of distant
galaxies with colours reddened by redshift.

\begin{figure*}[ht]
\centering
\includegraphics[width=1\textwidth]{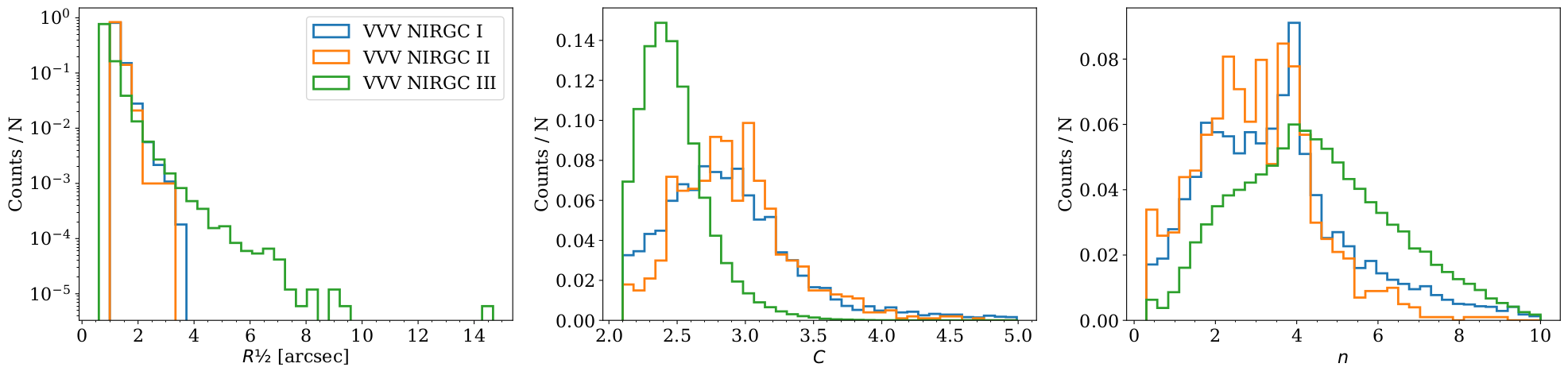}
\caption{Distributions of the morphological parameters: half-light radius, concentration index and Sersic index. The y-axis of the first distribution is in logarithmic scale.} 
\label{fig:histomorph}
\end{figure*}

Figure~\ref{fig:histomorph} shows the distribution of morphological parameters such as the half-light radius, concentration and Sersic indexes. The southern Galactic disc regions of VVV NIRGC III have lower interstellar extinctions, with the galaxies being both larger as well as fainter than those from the other two disc regions (VVV NIRGC I and II).  The half-light radius was not corrected for interstellar extinction. Consequently, it is expected that larger galaxies will be present in these regions. The concentration index distribution is similar for galaxies in VVV NIRGC I and II. For VVV NIRGC III, there is an increase in the number of galaxies with smaller C values, which correspond to late spirals \citep{Conselice2000}. Lower extinction might have allowed us to see the galaxy discs. 
The Sersic index distributions show different behaviours with peaks close to $n = 4$ for VVV NIRGC I and III, representing the de Vaucouleurs surface brightness profile for early-type galaxies. 

Table~\ref{tab:statistics} shows the median values of the extinction-corrected K${_s}$ magnitudes and colours, the morphological parameters R$_{1/2}$, $C$, $\epsilon$ and $n$ for the total number of galaxies in the VVV NIRGC III,  and the galaxies also found in the WISE survey. 
Comparing these results with those in Table 3 in \cite{Baravalle2021} for VVV NIRGC I, we are here reaching one magnitude fainter, from 14.24 to 15.19 mag in the $K^0_{s}$ passband. The Sersic index $n$ peaks around 4 in this work, which is different for VVV NIRGC I with some peaks in the range of 1.8 to 4.2 and a median value of 3.34. The R$_{1/2}$ values are comparable within the uncertainties reaching up to 3.40 arcsec for VVV NIRGC I and 14.68 arcsec for VVV NIRGC III. The other parameters and the standard deviations are comparable in general.

\begin{figure*}[ht]
\centering
\includegraphics[width=0.28\textwidth]{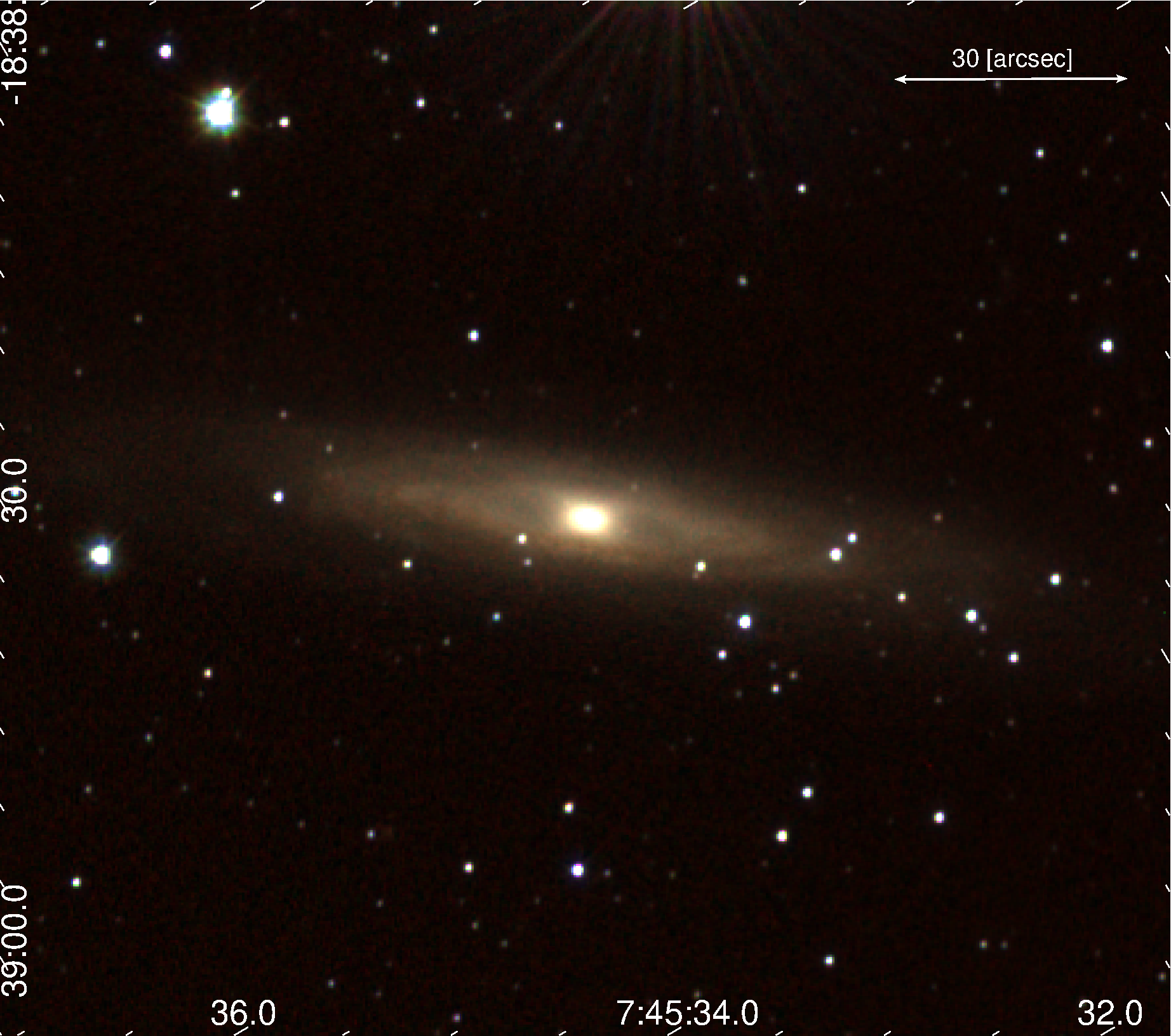}
\includegraphics[width=0.28\textwidth]{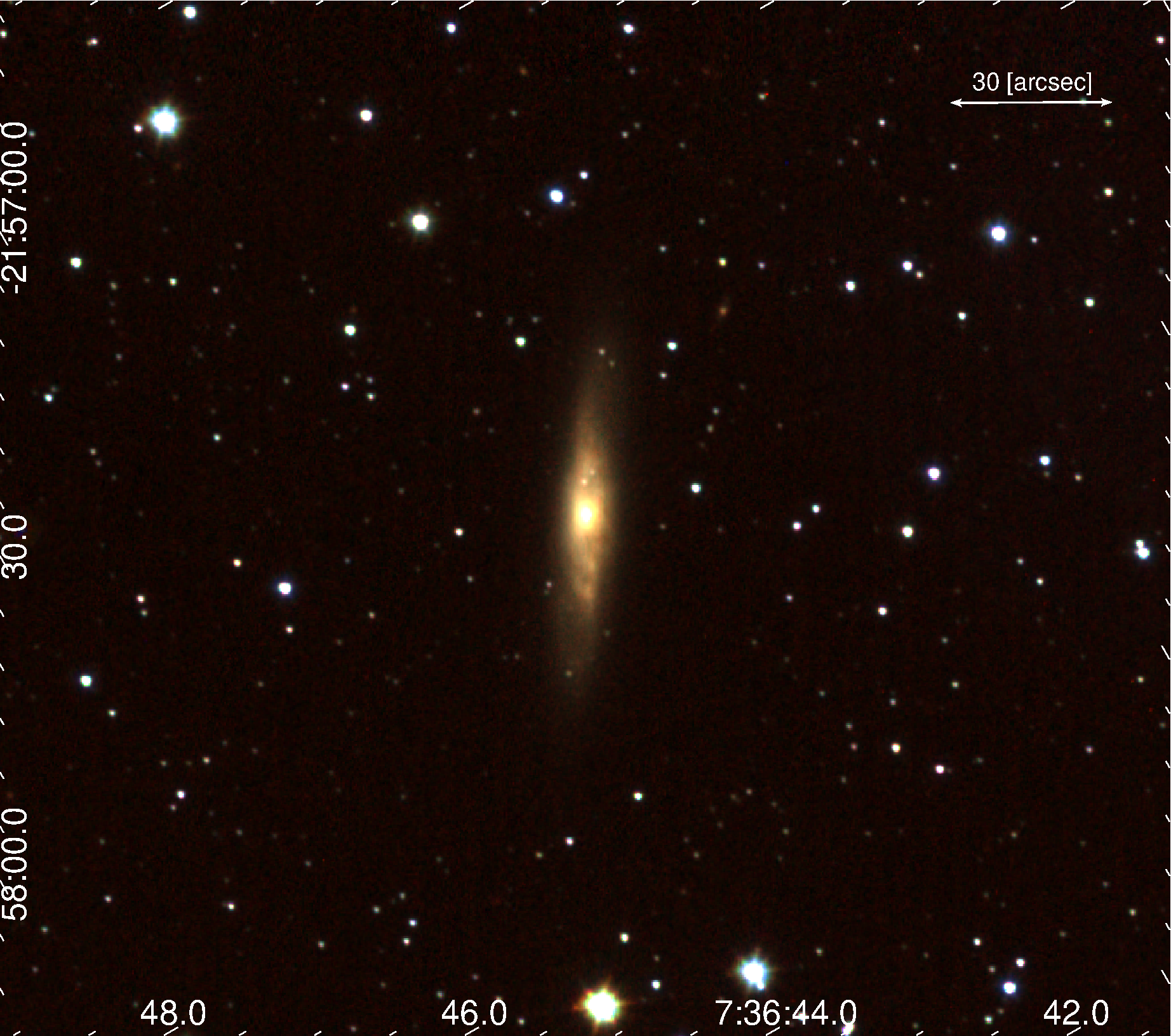}\\
\includegraphics[width=0.28\textwidth]{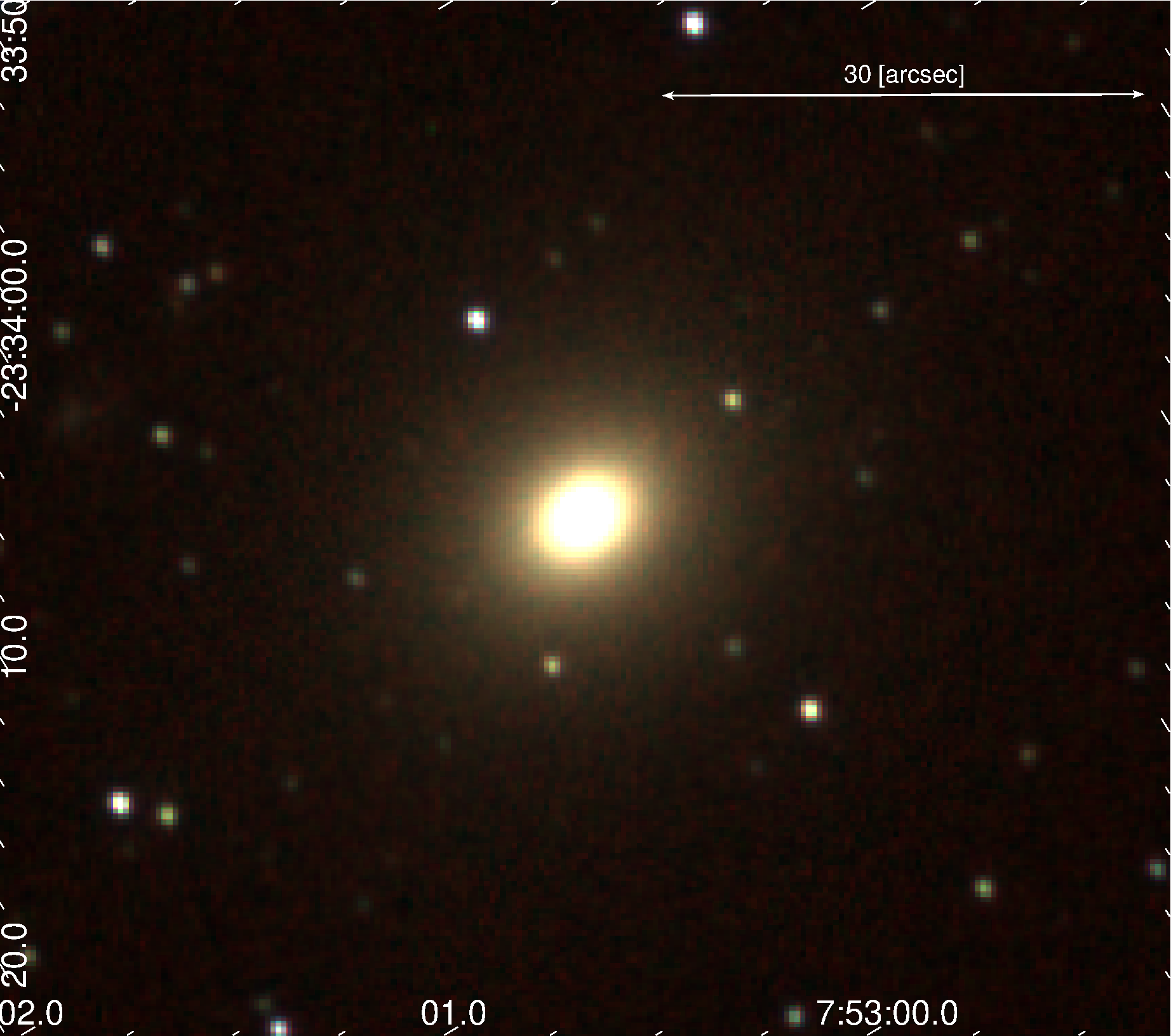}
\includegraphics[width=0.28\textwidth]{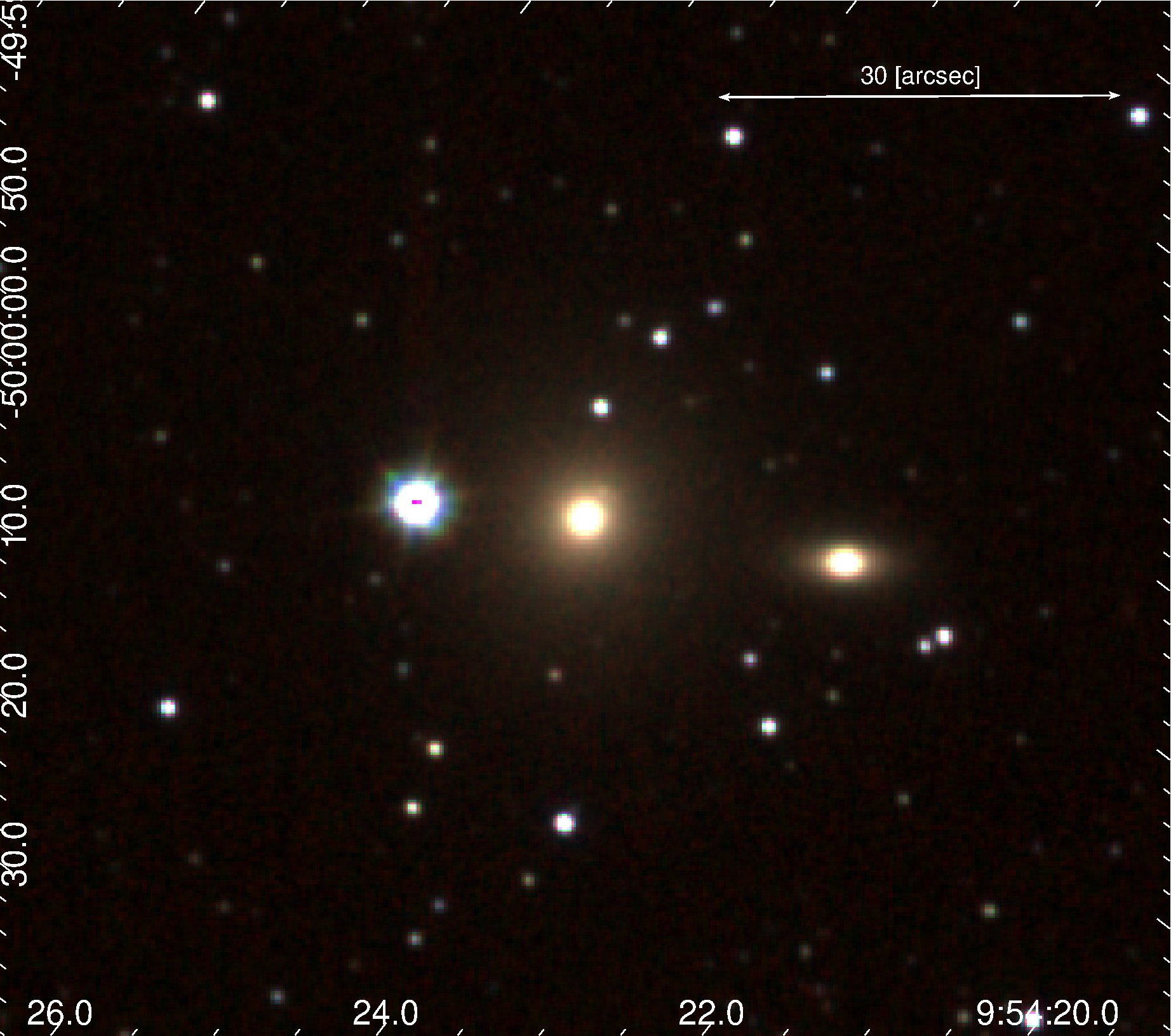}\\
\caption{Examples of the largest galaxies found in VVV NIRGC III. Upper panels show spiral galaxies and bottom panels, early-types. North is up and east is to the left.}
\label{fig:galaxies}
\end{figure*}

Figure~\ref{fig:galaxies} shows VVVX colour composed images of some examples of the VVV NIRGC III galaxies. The lower stellar density in these regions can be observed by comparing them with the innermost parts of the Galactic disc in VVV NIRGC I \citep{Baravalle2021}. Also, the lower interstellar extinction values allow us to see examples of spiral galaxies with faint surface brightness.

\begin{table}
\caption{Median photometric and structural parameters of the galaxies in VVV NIRGC III.}
\begin{tabular}{lcc}
\hline
\hline
Parameter & VVV NIRGC III &  with WISE data \\
 & 167,559 &  90,125  \\
\hline
\hline
$K^0_s$ (mag)       &  15.19 $\pm$ 0.89 &  15.01 $\pm$  0.96 \\
$(J - K_s)^0$ (mag) &   1.34 $\pm$ 0.23 &   1.33 $\pm$  0.23 \\
$(H - K_s)^0$ (mag) &   0.36 $\pm$ 0.15 &   0.35 $\pm$  0.14 \\
$R_{1/2}$ (arcsec)  &   0.80 $\pm$ 0.36 &   0.85 $\pm$  0.40 \\
C                   &   2.44 $\pm$ 0.25 &   2.49 $\pm$  0.27 \\
$\epsilon$          &   0.12 $\pm$ 0.15 &   0.25 $\pm$  0.22 \\
n                   &   4.31 $\pm$ 1.91 &   4.22 $\pm$  1.88 \\
\hline
\end{tabular}
\label{tab:statistics}
\end{table}

 \section{The galaxy distribution in the southern Galactic disc}
 \label{sec:distribution}

\begin{figure*}[ht]
\centering
\includegraphics[width=1\textwidth]{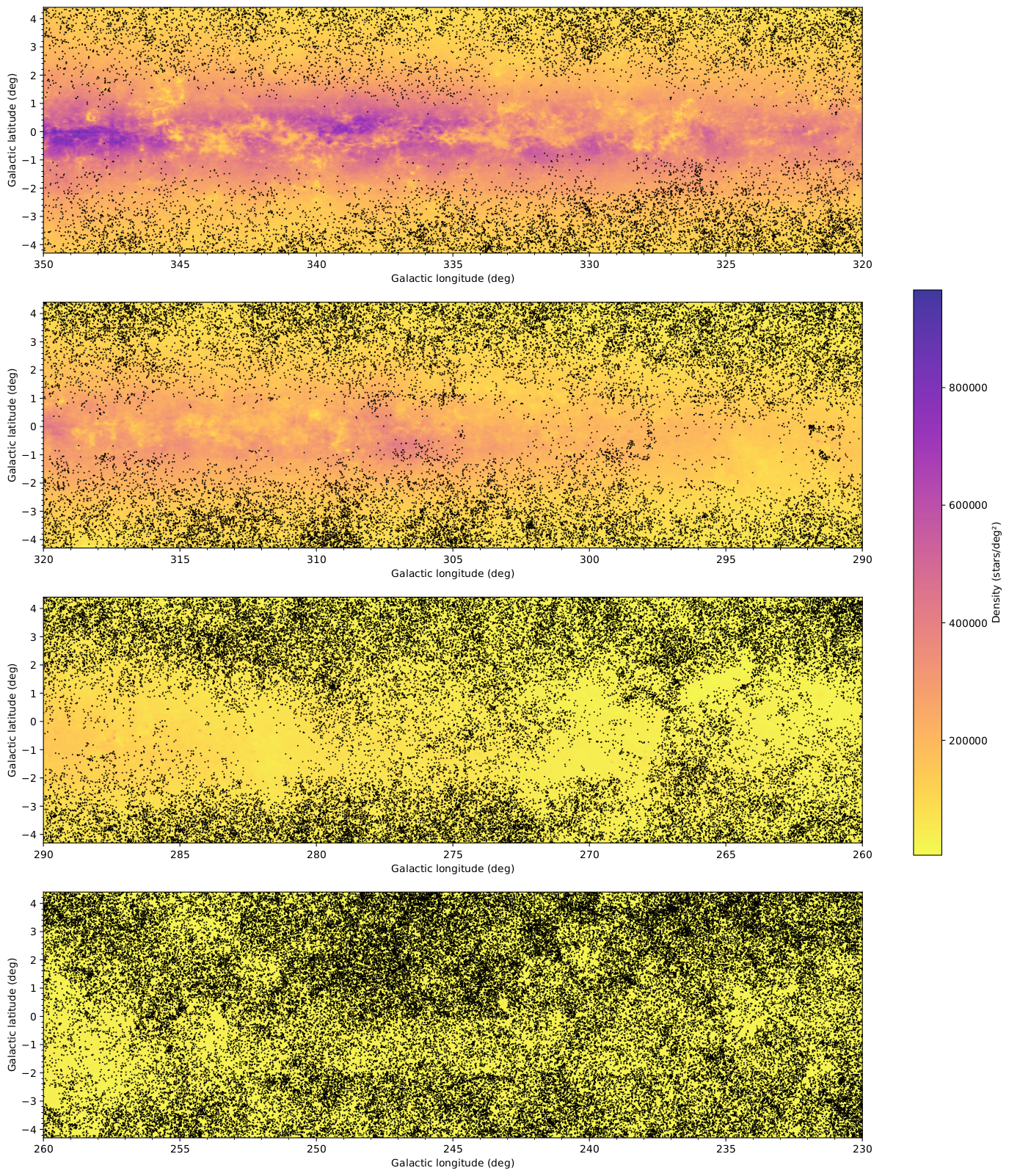}
\caption{Distribution of the galaxies in the southern Galactic disc using the VVV and VVVX surveys superimposed with the stellar density map with limiting magnitude of $K_s =$ 16 mag by Alonso-García et al. (in prep.).
The colour bar shows, on linear scale, the variation of the stellar density.
The four panels correspond to Galactic longitudes between 350$^\circ$ to 320$^\circ$, 320$^\circ$ to 290$^\circ$, 290$^\circ$ to 260$^\circ$, and 260$^\circ$ to 230$^\circ$. The galaxies from the VVV NIRGC I and VVV NIRGC III catalogues are both represented by black dots. }
\label{fig:distribution}
\end{figure*}

Figure~\ref{fig:distribution} shows the distribution of galaxies in the southern Galactic disc using the VVV NIRGC I and III catalogues with the VVV and VVVX surveys,
respectively, superimposed with the stellar density map from the PSF catalogues by Alonso-García et al. (in prep.). The galaxies from both catalogues are represented with black dots. In regions of higher interstellar extinction and stellar density, at Galactic longitudes between 350$^\circ$ to 320$^\circ$, it is almost impossible to detect galaxies. The distribution of galaxies increases markedly as we move away from the Galactic centre, and this is most prominent for Galactic longitudes between 260$^\circ$ to 230$^\circ$.  

The VVV NIRGC catalogues serve to complement the 2MASX catalogue in these star crowded regions of the Galactic disc where the separation of stars and galaxies poses a significant challenge.
The catalogue extends to $K^0_s = 16$ mag reaching more than 2 magnitudes deeper than 2MASX. 
Behind the southern Galactic disc, there is the Norma Cluster (ACO 3627; \citealt{Abell1989}), the most massive and rich cluster in the Great Attractor overdensity \citep{Kraan1996,Woudt1998}, located at  
$(l, b, z)$ = (325$^{\circ}$, -7$^{\circ}$, 0.016)  \citep{Skelton2009}. Additionally, the Norma Supercluster \citep{Woudt2001}, garnered significant attention in recent studies, particularly within the Galactic longitude range of 302$^{\circ}$ to 332$^{\circ}$, in close proximity to the Great Attractor 
\citep{Steyn2024}. This region has received comparatively less attention than other large-scale structures, primarily due to its location at low Galactic latitudes. However, due to the prominence of the Norma Supercluster and its importance in the dynamics of the local Universe, it is a crucial structure to study. 
The Vela region ($260^{\circ} < l < 290^{\circ}$, $-2^{\circ} < b < 1^{\circ}$) that includes the Vela Supercluster is also present in the figure. Complementary to our catalogue, in this region far from the Galactic disc, \cite{Hatamkhani2023, Hatamkhani2024} studied galaxy clusters. Other interesting structures are Perseus-Pisces at z $\approx$ 0.02 \citep{Focardi1984}, Laniakea at z $\approx$ 0.025 \citep{Tully2014}, and the Bootes and Sculptor voids at z $\approx$ 0.04 \citep{Kirshner1987}. Our catalogue's depth would allow for the study of distant galaxies at the Shapley Supercluster at z $\approx$ 0.05 \citep{Quintana1995}, thereby facilitating the investigation of the controversial Bulk Flow in the local Universe. Our work expands the mapping of these structures by including a larger number of galaxies. 


\section{Summary and final comments}\label{sec:conclusions}

In this work, we present the VVV NIRGC III catalogue using the VVVX survey of the southern Galactic disc (230$^{\circ}$ $<l<$ 350$^{\circ}$), which comprises the photometric and morphological properties of 167,559 galaxies having joint IS-CNN and PS-XGBoost probabilities $\geq$ 0.6, or 0.5 $\leq$ IS-CNN probabilities $\leq$ 0.6 and PS-XGBoost probabilities $\geq$ 0.8. 
The probabilities are the result of machine learning algorithms developed in \cite{Daza2023}, based on CNN using the $J$, $H$ and $K_{s}$ images and XGBoost using photometric and morphological parameters of the catalogue. 
Our catalogue has a contamination rate due to misclassifications of about 10\% and a completeness of 78\%. A total of 16,998 galaxies were visually confirmed, constituting 10\% of the total sample. Without duplicates, we have 23,333 confirmed galaxies which represents 14\% of the total catalogue, including 6,058 2MASX galaxies and 493 galaxies with spectroscopic redshifts and 3,670 galaxies with photometric redshifts. Upon request, it is available the larger catalogue of all NIR extragalactic candidate sources, with a total of 692,694 entries. This catalogue has no cuts in IS-CNN or PS-XGBoost probabilities, and only relies on morphological parameters, Gaia cross-matches and NIR colours to select sources (see \S 3 for details). 

This work contributes to the search for galaxies in the southern Galactic disc, complementing the 5,554 galaxies visually confirmed from VVV NIRGC I \citep{Baravalle2021} and resulting in a total of 173,113 galaxies in this part of the ZoA. We present a deep galaxy map down to $K^0_s = 16$ mag 
in the region $230^{\circ} \le l \le 350^{\circ}$ and $|b| \le 4.5^{\circ}$. The limiting magnitude is fainter than in previously studied regions (VVV NIRGC I and II) in the Galactic disc, mainly due to the smaller interstellar extinctions. 
The more prominent large-scale structures that still need further studies are the Norma and Vela Superclusters.

The Zone of Avoidance is beginning to reveal the distribution of galaxies obscured by the Galactic disc. At lower redshifts, the presence of both faint and bright galaxies is of significance, as they contribute to delineate the structure of the local Universe. This offers crucial insights into its role and significance within the broader context of models used to understand the Universe. 
As we have photometric and morphological parameters of the galaxies, the subsequent step is to focus efforts on redshift surveys to measure radial velocities. Spectroscopic observations may provide confirmation of the extragalactic nature of the objects presented in this new catalogue. 
Radial velocities can be obtained using the K-band Multi Object Spectrograph (KMOS, \citealt{Sharples2013}) at the ESO Very Large Telescope and the NIR imaging spectrograph
FLAMINGOS-2 \citep{Eikenberry2006} at Gemini South. 
It is important to note that HI surveys such as MeerKAT 
\citep{Goedhart2024} and WALLABY \citep{Koribalski2020, Westmeier2022}, can also provide radial velocities.   
 
New large-scale surveys, such as the Vera C.~Rubin Observatory Legacy Survey of Space
and Time \citep{LSST2009} and the upcoming Nancy Grace Roman Space Telescope \citep{Paladini2023}, should provide us with a substantial quantity of data at different wavelengths. Increased synergy with these surveys would help us to understand the structure and distribution of galaxies in the ZoA, thereby enriching our comprehension of the local Universe.

\begin{acknowledgements}

We thank the anonymous referee for useful comments and suggestions that have improved this paper.
This research was partially supported by Consejo de Investigaciones Cient\'ificas y T\'ecnicas (CONICET) and Secretar\'ia de Ciencia y T\'ecnica de la Universidad Nacional de C\'ordoba (SeCyT). We made use of the cross-match service provided by CDS, Strasbourg.
The authors gratefully acknowledge data from the ESO Public Survey program IDs 179.B-2002 and 198.B-2004 taken with the VISTA telescope, and products from the Cambridge Astronomical Survey Unit (CASU). 
M.V.A. would like to thank Drs. J. L. Sersic, L. A. Nicolaci da Costa and P. S. de Souza Pellegrini for their guidance during her early years in astronomy.
J.L.N.C. is grateful to the Universidad de La Serena for providing the academic environment and support that allowed the development of this research. This work was conducted without specific financial support, a fact mentioned here simply for clarity. 
M.S. acknowledges support from ANID’s FONDECYT Regular grant \#1251401. 
I.V.D.-P. acknowledges funding by NASA under the CRESST II program. D.M. gratefully acknowledges support provided by the ANID BASAL projects ACE210002 and FB210003 and by Fondecyt Project No. 1220724. J.A.-G. acknowledges support from ANID – Millennium Science Initiative Program – ICN12\_009 awarded to the Millennium Institute of Astrophysics MAS. R.K.S. acknowledges the support of CNPq/Brazil through projects 308298/2022- 5 and 421034/2023-8.
 We would also like to express our gratitude to C. Ragone Figueroa for her generous assistance.  
\end{acknowledgements}

\bibliographystyle{aa} 
\bibliography{aa52197}

\begin{thebibliography}{78}
\expandafter\ifx\csname natexlab\endcsname\relax\def\natexlab#1{#1}\fi

\bibitem[{{Abell} {et~al.}(1989){Abell}, {Corwin}, \& {Olowin}}]{Abell1989}
{Abell}, G.~O., {Corwin}, Harold~G., J., \& {Olowin}, R.~P. 1989, \apjs, 70, 1

\bibitem[{{Am{\^o}res} {et~al.}(2012){Am{\^o}res}, {Sodr{\'e}}, {Minniti},
  {Alonso}, {Padilla}, {Gurovich}, {Arsenijevic}, {Tollerud},
  {Rodr{\'{\i}}guez-Ardila}, {D{\'{\i}}az Tello}, \& {Lucas}}]{Amores2012}
{Am{\^o}res}, E.~B., {Sodr{\'e}}, L., {Minniti}, D., {et~al.} 2012, \aj, 144,
  127

\bibitem[{{Asmus} {et~al.}(2020){Asmus}, {Greenwell}, {Gandhi}, {Boorman},
  {Aird}, {Alexander}, {Assef}, {Baldi}, {Davies}, {H{\"o}nig}, {Ricci},
  {Rosario}, {Salvato}, {Shankar}, \& {Stern}}]{Asmus2020}
{Asmus}, D., {Greenwell}, C.~L., {Gandhi}, P., {et~al.} 2020, \mnras, 494, 1784

\bibitem[{{Baravalle} {et~al.}(2021){Baravalle}, {Alonso}, {Minniti}, {Nilo
  Castell{\'o}n}, {Soto}, {Valotto}, {Villal{\'o}n}, {Gra{\~n}a}, {Am{\^o}res},
  \& {Milla Castro}}]{Baravalle2021}
{Baravalle}, L.~D., {Alonso}, M.~V., {Minniti}, D., {et~al.} 2021, \mnras, 502,
  601

\bibitem[{{Baravalle} {et~al.}(2018){Baravalle}, {Alonso}, {Nilo
  Castell{\'o}n}, {Beam{\'{\i}}n}, \& {Minniti}}]{Baravalle2018}
{Baravalle}, L.~D., {Alonso}, M.~V., {Nilo Castell{\'o}n}, J.~L.,
  {Beam{\'{\i}}n}, J.~C., \& {Minniti}, D. 2018, \aj, 155, 46

\bibitem[{{Baravalle} {et~al.}(2023){Baravalle}, {Schmidt}, {Alonso}, {Pichel},
  {Minniti}, {Rodr{\'\i}guez-Kamenetzky}, {Masetti}, {Villalon}, {Smith}, \&
  {Lucas}}]{Baravalle2023}
{Baravalle}, L.~D., {Schmidt}, E.~O., {Alonso}, M.~V., {et~al.} 2023, \mnras,
  520, 5950

\bibitem[{{Bertin}(2011)}]{Bertin2011}
{Bertin}, E. 2011, in Astronomical Society of the Pacific Conference Series,
  Vol. 442, Astronomical Data Analysis Software and Systems XX, ed. I.~N.
  {Evans}, A.~{Accomazzi}, D.~J. {Mink}, \& A.~H. {Rots}, 435

\bibitem[{{Bertin} \& {Arnouts}(1996)}]{Bertin1996}
{Bertin}, E. \& {Arnouts}, S. 1996, \aaps, 117, 393

\bibitem[{{Bilicki} {et~al.}(2014){Bilicki}, {Jarrett}, {Peacock}, {Cluver}, \&
  {Steward}}]{Bilicki2014}
{Bilicki}, M., {Jarrett}, T.~H., {Peacock}, J.~A., {Cluver}, M.~E., \&
  {Steward}, L. 2014, \apjs, 210, 9

\bibitem[{{Catelan} {et~al.}(2011){Catelan}, {Minniti}, {Lucas},
  {Alonso-Garc{\'{\i}}a}, {Angeloni}, {Beam{\'{\i}}n}, {Bonatto}, {Borissova},
  {Contreras}, {Cross}, {D{\'e}k{\'a}any}, {Emerson}, {Eyheramendy}, {Geisler},
  {Gonz{\'a}lez-Solares}, {Helminiak}, {Hempel}, {Irwin}, {Ivanov},
  {Jord{\'a}n}, {Kerins}, {Kurtev}, {Mauro}, {Moni Bidin}, {Navarrete},
  {P{\'e}rez}, {Pichara}, {Read}, {Rejkuba}, {Saito}, {Sale}, \&
  {Toledo}}]{Catelan2011}
{Catelan}, M., {Minniti}, D., {Lucas}, P.~W., {et~al.} 2011, in RR Lyrae Stars,
  Metal-Poor Stars, and the Galaxy, ed. A.~{McWilliam}, Vol.~5, 145

\bibitem[{Chen \& Guestrin(2016)}]{Chen2016}
Chen, T. \& Guestrin, C. 2016, in Proceedings of the 22nd ACM SIGKDD
  International Conference on Knowledge Discovery and Data Mining, KDD '16 (New
  York, NY, USA: Association for Computing Machinery), 785–794

\bibitem[{Cheng {et~al.}(2021)Cheng, Conselice, Arag{\'{o} }n-Salamanca,
  Aguena, Allam, Andrade-Oliveira, Annis, Bluck, Brooks, Burke, Kind,
  Carretero, Choi, Costanzi, da~Costa, Pereira, Vicente, Diehl, Drlica-Wagner,
  Eckert, Everett, Evrard, Ferrero, Fosalba, Frieman, Garc{\'{\i}}a-Bellido,
  Gerdes, Giannantonio, Gruen, Gruendl, Gschwend, Gutierrez, Hinton, Hollowood,
  Honscheid, James, Krause, Kuehn, Kuropatkin, Lahav, Maia, March, Menanteau,
  Miquel, Morgan, Paz-Chinch{\'{o}}n, Pieres, Malag{\'{o}}n, Roodman, Sanchez,
  Scarpine, Serrano, Sevilla-Noarbe, Smith, Soares-Santos, Suchyta, Swanson,
  Tarle, Thomas, \& To}]{Cheng_2021}
Cheng, T.-Y., Conselice, C.~J., Arag{\'{o} }n-Salamanca, A., {et~al.} 2021,
  Monthly Notices of the Royal Astronomical Society, 507, 4425

\bibitem[{{Clocchiatti} {et~al.}(2024){Clocchiatti}, {Rodr{\'\i}guez},
  {{\'O}rdenes Morales}, \& {Cuevas-Tapia}}]{Clocchiatti2024}
{Clocchiatti}, A., {Rodr{\'\i}guez}, {\'O}., {{\'O}rdenes Morales}, A., \&
  {Cuevas-Tapia}, B. 2024, \apj, 971, 19

\bibitem[{{Coldwell} {et~al.}(2014){Coldwell}, {Alonso}, {Duplancic}, {Hempel},
  {Ivanov}, \& {Minniti}}]{Coldwell2014}
{Coldwell}, G., {Alonso}, S., {Duplancic}, F., {et~al.} 2014, \aap, 569, A49

\bibitem[{{Conselice} {et~al.}(2000){Conselice}, {Bershady}, \&
  {Jangren}}]{Conselice2000}
{Conselice}, C.~J., {Bershady}, M.~A., \& {Jangren}, A. 2000, \apj, 529, 886

\bibitem[{{Cutri} {et~al.}(2003){Cutri}, {Skrutskie}, {van Dyk}, {Beichman},
  {Carpenter}, {Chester}, {Cambresy}, {Evans}, {Fowler}, {Gizis}, {Howard},
  {Huchra}, {Jarrett}, {Kopan}, {Kirkpatrick}, {Light}, {Marsh}, {McCallon},
  {Schneider}, {Stiening}, {Sykes}, {Weinberg}, {Wheaton}, {Wheelock}, \&
  {Zacarias}}]{Cutri2003}
{Cutri}, R.~M., {Skrutskie}, M.~F., {van Dyk}, S., {et~al.} 2003, VizieR Online
  Data Catalog, II/246

\bibitem[{{Daza-Perilla} {et~al.}(2023){Daza-Perilla}, {Sgr{\'o}}, {Baravalle},
  {Alonso}, {Villalon}, {Lares}, {Soto}, {Castell{\'o}n}, {Valotto},
  {Cort{\'e}s}, {Minniti}, \& {Hempel}}]{Daza2023}
{Daza-Perilla}, I.~V., {Sgr{\'o}}, M.~A., {Baravalle}, L.~D., {et~al.} 2023,
  \mnras, 524, 678

\bibitem[{de~Lima(2024)}]{delima_specz_compilation}
de~Lima, E. V.~R. 2024, {ErikVini/specz\_compilation: Southern Hemisphere
  Spectrocopic Redshift Compilation}

\bibitem[{{Donoso} {et~al.}(2024){Donoso}, {Pichel}, {Baravalle}, {Alonso},
  {Schmidt}, {Minniti}, {Masetti}, {Smith}, {Lucas}, {Villalon}, {Rovero}, \&
  {Coldwell}}]{Donoso2024}
{Donoso}, L.~G., {Pichel}, A., {Baravalle}, L.~D., {et~al.} 2024, \mnras, 529,
  1019

\bibitem[{{Duplancic} {et~al.}(2024){Duplancic}, {Alonso}, {Coldwell},
  {Galdeano}, {Minniti}, {Fernandez}, {Mesa}, {Perez}, {Pereyra}, \&
  {Pavesich}}]{Duplancic2024}
{Duplancic}, F., {Alonso}, S., {Coldwell}, G., {et~al.} 2024, \aap, 682, A153

\bibitem[{{Eikenberry} {et~al.}(2006){Eikenberry}, {Elston}, {Raines},
  {Julian}, {Hanna}, {Hon}, {Julian}, {Bandyopadhyay}, {Bennett}, {Bessoff},
  {Branch}, {Corley}, {Eriksen}, {Frommeyer}, {Gonzalez}, {Herlevich},
  {Marin-Franch}, {Marti}, {Murphey}, {Rashkin}, {Warner}, {Leckie},
  {Gardhouse}, {Fletcher}, {Dunn}, {Wooff}, \& {Hardy}}]{Eikenberry2006}
{Eikenberry}, S., {Elston}, R., {Raines}, S.~N., {et~al.} 2006, in Society of
  Photo-Optical Instrumentation Engineers (SPIE) Conference Series, Vol. 6269,
  Ground-based and Airborne Instrumentation for Astronomy, ed. I.~S. {McLean}
  \& M.~{Iye}, 626917

\bibitem[{{Focardi} {et~al.}(1984){Focardi}, {Marano}, \&
  {Vettolani}}]{Focardi1984}
{Focardi}, P., {Marano}, B., \& {Vettolani}, G. 1984, \aap, 136, 178

\bibitem[{{Freeman} {et~al.}(1977){Freeman}, {Karlsson}, {Lynga}, {Burrell},
  {van Woerden}, {Goss}, \& {Mebold}}]{Freeman1977}
{Freeman}, K.~C., {Karlsson}, B., {Lynga}, G., {et~al.} 1977, \aap, 55, 445

\bibitem[{{Gaia Collaboration}(2020)}]{2020yCat.1350....0G}
{Gaia Collaboration}. 2020, VizieR Online Data Catalog, I/350

\bibitem[{{Gaia Collaboration} {et~al.}(2021){Gaia Collaboration}, {Brown},
  {Vallenari}, {Prusti}, {de Bruijne}, {Babusiaux}, {Biermann}, {Creevey},
  {Evans}, {Eyer}, {Hutton}, {Jansen}, {Jordi}, {Klioner}, {Lammers},
  {Lindegren}, {Luri}, {Mignard}, {Panem}, {Pourbaix}, {Randich}, {Sartoretti},
  {Soubiran}, {Walton}, {Arenou}, {Bailer-Jones}, {Bastian}, {Cropper},
  {Drimmel}, {Katz}, {Lattanzi}, {van Leeuwen}, {Bakker}, {Cacciari},
  {Casta{\~n}eda}, {De Angeli}, {Ducourant}, {Fabricius}, {Fouesneau},
  {Fr{\'e}mat}, {Guerra}, {Guerrier}, {Guiraud}, {Jean-Antoine Piccolo},
  {Masana}, {Messineo}, {Mowlavi}, {Nicolas}, {Nienartowicz}, {Pailler},
  {Panuzzo}, {Riclet}, {Roux}, {Seabroke}, {Sordo}, {Tanga}, {Th{\'e}venin},
  {Gracia-Abril}, {Portell}, {Teyssier}, {Altmann}, {Andrae}, {Bellas-Velidis},
  {Benson}, {Berthier}, {Blomme}, {Brugaletta}, {Burgess}, {Busso}, {Carry},
  {Cellino}, {Cheek}, {Clementini}, {Damerdji}, {Davidson}, {Delchambre},
  {Dell'Oro}, {Fern{\'a}ndez-Hern{\'a}ndez}, {Galluccio}, {Garc{\'\i}a-Lario},
  {Garcia-Reinaldos}, {Gonz{\'a}lez-N{\'u}{\~n}ez}, {Gosset}, {Haigron},
  {Halbwachs}, {Hambly}, {Harrison}, {Hatzidimitriou}, {Heiter},
  {Hern{\'a}ndez}, {Hestroffer}, {Hodgkin}, {Holl}, {Jan{\ss}en}, {Jevardat de
  Fombelle}, {Jordan}, {Krone-Martins}, {Lanzafame}, {L{\"o}ffler}, {Lorca},
  {Manteiga}, {Marchal}, {Marrese}, {Moitinho}, {Mora}, {Muinonen}, {Osborne},
  {Pancino}, {Pauwels}, {Petit}, {Recio-Blanco}, {Richards}, {Riello},
  {Rimoldini}, {Robin}, {Roegiers}, {Rybizki}, {Sarro}, {Siopis}, {Smith},
  {Sozzetti}, {Ulla}, {Utrilla}, {van Leeuwen}, {van Reeven}, {Abbas}, {Abreu
  Aramburu}, {Accart}, {Aerts}, {Aguado}, {Ajaj}, {Altavilla}, {{\'A}lvarez},
  {{\'A}lvarez Cid-Fuentes}, {Alves}, {Anderson}, {Anglada Varela}, {Antoja},
  {Audard}, {Baines}, {Baker}, {Balaguer-N{\'u}{\~n}ez}, {Balbinot}, {Balog},
  {Barache}, {Barbato}, {Barros}, {Barstow}, {Bartolom{\'e}}, {Bassilana},
  {Bauchet}, {Baudesson-Stella}, {Becciani}, {Bellazzini}, {Bernet}, {Bertone},
  {Bianchi}, {Blanco-Cuaresma}, {Boch}, {Bombrun}, {Bossini}, {Bouquillon},
  {Bragaglia}, {Bramante}, {Breedt}, {Bressan}, {Brouillet}, {Bucciarelli},
  {Burlacu}, {Busonero}, {Butkevich}, {Buzzi}, {Caffau}, {Cancelliere},
  {C{\'a}novas}, {Cantat-Gaudin}, {Carballo}, {Carlucci}, {Carnerero},
  {Carrasco}, {Casamiquela}, {Castellani}, {Castro-Ginard}, {Castro Sampol},
  {Chaoul}, {Charlot}, {Chemin}, {Chiavassa}, {Cioni}, {Comoretto}, {Cooper},
  {Cornez}, {Cowell}, {Crifo}, {Crosta}, {Crowley}, {Dafonte}, {Dapergolas},
  {David}, {David}, {de Laverny}, {De Luise}, {De March}, {De Ridder}, {de
  Souza}, {de Teodoro}, {de Torres}, {del Peloso}, {del Pozo}, {Delbo},
  {Delgado}, {Delgado}, {Delisle}, {Di Matteo}, {Diakite}, {Diener},
  {Distefano}, {Dolding}, {Eappachen}, {Edvardsson}, {Enke}, {Esquej}, {Fabre},
  {Fabrizio}, {Faigler}, {Fedorets}, {Fernique}, {Fienga}, {Figueras},
  {Fouron}, {Fragkoudi}, {Fraile}, {Franke}, {Gai}, {Garabato},
  {Garcia-Gutierrez}, {Garc{\'\i}a-Torres}, {Garofalo}, {Gavras}, {Gerlach},
  {Geyer}, {Giacobbe}, {Gilmore}, {Girona}, {Giuffrida}, {Gomel}, {Gomez},
  {Gonzalez-Santamaria}, {Gonz{\'a}lez-Vidal}, {Granvik},
  {Guti{\'e}rrez-S{\'a}nchez}, {Guy}, {Hauser}, {Haywood}, {Helmi}, {Hidalgo},
  {Hilger}, {H{\l}adczuk}, {Hobbs}, {Holland}, {Huckle}, {Jasniewicz},
  {Jonker}, {Juaristi Campillo}, {Julbe}, {Karbevska}, {Kervella}, {Khanna},
  {Kochoska}, {Kontizas}, {Kordopatis}, {Korn}, {Kostrzewa-Rutkowska},
  {Kruszy{\'n}ska}, {Lambert}, {Lanza}, {Lasne}, {Le Campion}, {Le Fustec},
  {Lebreton}, {Lebzelter}, {Leccia}, {Leclerc}, {Lecoeur-Taibi}, {Liao},
  {Licata}, {Lindstr{\o}m}, {Lister}, {Livanou}, {Lobel}, {Madrero Pardo},
  {Managau}, {Mann}, {Marchant}, {Marconi}, {Marcos Santos}, {Marinoni},
  {Marocco}, {Marshall}, {Martin Polo}, {Mart{\'\i}n-Fleitas}, {Masip},
  {Massari}, {Mastrobuono-Battisti}, {Mazeh}, {McMillan}, {Messina},
  {Michalik}, {Millar}, {Mints}, {Molina}, {Molinaro}, {Moln{\'a}r},
  {Montegriffo}, {Mor}, {Morbidelli}, {Morel}, {Morris}, {Mulone}, {Munoz},
  {Muraveva}, {Murphy}, {Musella}, {Noval}, {Ord{\'e}novic}, {Orr{\`u}},
  {Osinde}, {Pagani}, {Pagano}, {Palaversa}, {Palicio}, {Panahi}, {Pawlak},
  {Pe{\~n}alosa Esteller}, {Penttil{\"a}}, {Piersimoni}, {Pineau}, {Plachy},
  {Plum}, {Poggio}, {Poretti}, {Poujoulet}, {Pr{\v{s}}a}, {Pulone}, {Racero},
  {Ragaini}, {Rainer}, {Raiteri}, {Rambaux}, {Ramos}, {Ramos-Lerate}, {Re
  Fiorentin}, {Regibo}, {Reyl{\'e}}, {Ripepi}, {Riva}, {Rixon}, {Robichon},
  {Robin}, {Roelens}, {Rohrbasser}, {Romero-G{\'o}mez}, {Rowell}, {Royer},
  {Rybicki}, {Sadowski}, {Sagrist{\`a} Sell{\'e}s}, {Sahlmann}, {Salgado},
  {Salguero}, {Samaras}, {Sanchez Gimenez}, {Sanna}, {Santove{\~n}a},
  {Sarasso}, {Schultheis}, {Sciacca}, {Segol}, {Segovia}, {S{\'e}gransan},
  {Semeux}, {Shahaf}, {Siddiqui}, {Siebert}, {Siltala}, {Slezak}, {Smart},
  {Solano}, {Solitro}, {Souami}, {Souchay}, {Spagna}, {Spoto}, {Steele},
  {Steidelm{\"u}ller}, {Stephenson}, {S{\"u}veges}, {Szabados}, {Szegedi-Elek},
  {Taris}, {Tauran}, {Taylor}, {Teixeira}, {Thuillot}, {Tonello}, {Torra},
  {Torra}, {Turon}, {Unger}, {Vaillant}, {van Dillen}, {Vanel}, {Vecchiato},
  {Viala}, {Vicente}, {Voutsinas}, {Weiler}, {Wevers}, {Wyrzykowski}, {Yoldas},
  {Yvard}, {Zhao}, {Zorec}, {Zucker}, {Zurbach}, \&
  {Zwitter}}]{2021A&A...650C...3G}
{Gaia Collaboration}, {Brown}, A.~G.~A., {Vallenari}, A., {et~al.} 2021, \aap,
  650, C3

\bibitem[{{Galdeano} {et~al.}(2022){Galdeano}, {Coldwell}, {Duplancic},
  {Alonso}, {Pereyra}, {Minniti}, {Zelada Bacigalupo}, {Valotto}, {Baravalle},
  {Alonso}, \& {Nilo Castell{\'o}n}}]{Galdeano2022}
{Galdeano}, D., {Coldwell}, G., {Duplancic}, F., {et~al.} 2022, \aap, 663, A158

\bibitem[{{Galdeano} {et~al.}(2023){Galdeano}, {Ferrero}, {Coldwell},
  {Duplancic}, {Alonso}, {Riffel}, \& {Minniti}}]{Galdeano2023}
{Galdeano}, D., {Ferrero}, G.~A., {Coldwell}, G., {et~al.} 2023, \aap, 669, A7

\bibitem[{{Galdeano} {et~al.}(2021){Galdeano}, {Pereyra}, {Duplancic},
  {Coldwell}, {Alonso}, {Ruiz}, {Cora}, {Perez}, {Vega-Mart{\'\i}nez}, \&
  {Minniti}}]{Galdeano2021}
{Galdeano}, D., {Pereyra}, L., {Duplancic}, F., {et~al.} 2021, \aap, 646, A146

\bibitem[{{Goedhart} {et~al.}(2024){Goedhart}, {Cotton}, {Camilo}, {Thompson},
  {Umana}, {Bietenholz}, {Woudt}, {Anderson}, {Bordiu}, {Buckley}, {Buemi},
  {Bufano}, {Cavallaro}, {Chen}, {Chibueze}, {Egbo}, {Frank}, {Hoare},
  {Ingallinera}, {Irabor}, {Kraan-Korteweg}, {Kurapati}, {Leto}, {Loru},
  {Mutale}, {Obonyo}, {Plavin}, {Rajohnson}, {Rigby}, {Riggi}, {Seidu},
  {Serra}, {Smart}, {Stappers}, {Steyn}, {Surnis}, {Trigilio}, {Williams},
  {Abbott}, {Adam}, {Asad}, {Baloyi}, {Bauermeister}, {Bennet}, {Bester},
  {Botha}, {Brederode}, {Buchner}, {Burger}, {Cheetham}, {Cloete}, {de
  Villiers}, {de Villiers}, {du Toit}, {Esterhuyse}, {Fanaroff}, {Fourie},
  {Gamatham}, {Gatsi}, {Geyer}, {Gouws}, {Gumede}, {Heywood}, {Hokwana},
  {Hoosen}, {Horn}, {Horrell}, {Hugo}, {Isaacson}, {J{\'o}zsa}, {Jonas},
  {Jordaan}, {Joubert}, {Julie}, {Kapp}, {Kriek}, {Kriel}, {Krishnan}, {Kusel},
  {Legodi}, {Lehmensiek}, {Lord}, {Macfarlane}, {Magnus}, {Magozore}, {Main},
  {Malan}, {Manley}, {Marais}, {Maree}, {Martens}, {Maruping}, {McAlpine},
  {Merry}, {Mgodeli}, {Millenaar}, {Mokone}, {Monama}, {New}, {Ngcebetsha},
  {Ngoasheng}, {Nicolson}, {Ockards}, {Oozeer}, {Passmoor}, {Patel},
  {Peens-Hough}, {Perkins}, {Ramaila}, {Ratcliffe}, {Renil}, {Richter},
  {Salie}, {Sambu}, {Schollar}, {Schwardt}, {Schwartz}, {Serylak}, {Siebrits},
  {Sirothia}, {Slabber}, {Smirnov}, {Tiplady}, {van Balla}, {van der Byl}, {Van
  Tonder}, {Venter}, {Venter}, {Welz}, \& {Williams}}]{Goedhart2024}
{Goedhart}, S., {Cotton}, W.~D., {Camilo}, F., {et~al.} 2024, \mnras, 531, 649

\bibitem[{{Green}(2018)}]{dustmaps}
{Green}, G. 2018, The Journal of Open Source Software, 3, 695

\bibitem[{{Hatamkhani} {et~al.}(2023){Hatamkhani}, {Kraan-Korteweg}, {Blyth},
  {Said}, \& {Elagali}}]{Hatamkhani2023}
{Hatamkhani}, N., {Kraan-Korteweg}, R.~C., {Blyth}, S.~L., {Said}, K., \&
  {Elagali}, A. 2023, \mnras, 522, 2223

\bibitem[{{Hatamkhani} {et~al.}(2024){Hatamkhani}, {Kraan-Korteweg}, {Blyth},
  \& {Skelton}}]{Hatamkhani2024}
{Hatamkhani}, N., {Kraan-Korteweg}, R.~C., {Blyth}, S.~L., \& {Skelton}, R.~E.
  2024, \apj, 972, 57

\bibitem[{{Henning}(1992)}]{Henning1992}
{Henning}, P.~A. 1992, \apjs, 78, 365

\bibitem[{{Huchra} {et~al.}(2005){Huchra}, {Jarrett}, {Skrutskie}, {Cutri},
  {Schneider}, {Macri}, {Steining}, {Mader}, {Martimbeau}, \&
  {George}}]{Huchra2005}
{Huchra}, J., {Jarrett}, T., {Skrutskie}, M., {et~al.} 2005, in Astronomical
  Society of the Pacific Conference Series, Vol. 329, Nearby Large-Scale
  Structures and the Zone of Avoidance, ed. A.~P. {Fairall} \& P.~A. {Woudt},
  135

\bibitem[{{Huchra} {et~al.}(2012){Huchra}, {Macri}, {Masters}, {Jarrett},
  {Berlind}, {Calkins}, {Crook}, {Cutri}, {Erdo{\v g}du}, {Falco}, {George},
  {Hutcheson}, {Lahav}, {Mader}, {Mink}, {Martimbeau}, {Schneider},
  {Skrutskie}, {Tokarz}, \& {Westover}}]{Huchra2012}
{Huchra}, J.~P., {Macri}, L.~M., {Masters}, K.~L., {et~al.} 2012, Astrophys. J.
  Suppl., 199, 26

\bibitem[{{Jarrett}(2004)}]{Jarrett2004}
{Jarrett}, T. 2004, \pasa, 21, 396

\bibitem[{{Jarrett} {et~al.}(2000){Jarrett}, {Chester}, {Cutri}, \&
  {Schneider}}]{2000AJ....119.2498J}
{Jarrett}, T.~H., {Chester}, T., {Cutri}, R., \& {Schneider}, S. 2000, \aj,
  119, 2498

\bibitem[{{Kirshner} {et~al.}(1987){Kirshner}, {Oemler}, {Schechter}, \&
  {Shectman}}]{Kirshner1987}
{Kirshner}, R.~P., {Oemler}, Jr., A., {Schechter}, P.~L., \& {Shectman}, S.~A.
  1987, \apj, 314, 493

\bibitem[{{Komatsu} {et~al.}(2011){Komatsu}, {Smith}, {Dunkley}, {Bennett},
  {Gold}, {Hinshaw}, {Jarosik}, {Larson}, {Nolta}, {Page}, {Spergel},
  {Halpern}, {Hill}, {Kogut}, {Limon}, {Meyer}, {Odegard}, {Tucker}, {Weiland},
  {Wollack}, \& {Wright}}]{Komatsu2011}
{Komatsu}, E., {Smith}, K.~M., {Dunkley}, J., {et~al.} 2011, \apjs, 192, 18

\bibitem[{{Koribalski} {et~al.}(2020){Koribalski}, {Staveley-Smith},
  {Westmeier}, {Serra}, {Spekkens}, {Wong}, {Lee-Waddell}, {Lagos},
  {Obreschkow}, {Ryan-Weber}, {Zwaan}, {Kilborn}, {Bekiaris}, {Bekki},
  {Bigiel}, {Boselli}, {Bosma}, {Catinella}, {Chauhan}, {Cluver}, {Colless},
  {Courtois}, {Crain}, {de Blok}, {D{\'e}nes}, {Duffy}, {Elagali}, {Fluke},
  {For}, {Heald}, {Henning}, {Hess}, {Holwerda}, {Howlett}, {Jarrett}, {Jones},
  {Jones}, {J{\'o}zsa}, {Jurek}, {J{\"u}tte}, {Kamphuis}, {Karachentsev},
  {Kerp}, {Kleiner}, {Kraan-Korteweg}, {L{\'o}pez-S{\'a}nchez}, {Madrid},
  {Meyer}, {Mould}, {Murugeshan}, {Norris}, {Oh}, {Oosterloo}, {Popping},
  {Putman}, {Reynolds}, {Rhee}, {Robotham}, {Ryder}, {Schr{\"o}der}, {Shao},
  {Stevens}, {Taylor}, {van{\^A} der Hulst}, {Verdes-Montenegro}, {Wakker},
  {Wang}, {Whiting}, {Winkel}, \& {Wolf}}]{Koribalski2020}
{Koribalski}, B.~S., {Staveley-Smith}, L., {Westmeier}, T., {et~al.} 2020,
  \apss, 365, 118

\bibitem[{{Kourkchi} \& {Tully}(2017)}]{Kourkchi2017}
{Kourkchi}, E. \& {Tully}, R.~B. 2017, \apj, 843, 16

\bibitem[{{Kraan-Korteweg} {et~al.}(2018){Kraan-Korteweg}, {van Driel},
  {Schr{\"o}der}, {Ramatsoku}, \& {Henning}}]{Kraan2018}
{Kraan-Korteweg}, R.~C., {van Driel}, W., {Schr{\"o}der}, A.~C., {Ramatsoku},
  M., \& {Henning}, P.~A. 2018, \mnras, 481, 1262

\bibitem[{{Kraan-Korteweg} {et~al.}(1996){Kraan-Korteweg}, {Woudt}, {Cayatte},
  {Fairall}, {Balkowski}, \& {Henning}}]{Kraan1996}
{Kraan-Korteweg}, R.~C., {Woudt}, P.~A., {Cayatte}, V., {et~al.} 1996, \nat,
  379, 519

\bibitem[{{Kron}(1980)}]{Kron1980}
{Kron}, R.~G. 1980, Astrophys. J. Suppl., 43, 305

\bibitem[{{LSST Science Collaboration} {et~al.}(2009){LSST Science
  Collaboration}, {Abell}, {Allison}, {Anderson}, {Andrew}, {Angel}, {Armus},
  {Arnett}, {Asztalos}, {Axelrod}, {Bailey}, {Ballantyne}, {Bankert},
  {Barkhouse}, {Barr}, {Barrientos}, {Barth}, {Bartlett}, {Becker}, {Becla},
  {Beers}, {Bernstein}, {Biswas}, {Blanton}, {Bloom}, {Bochanski}, {Boeshaar},
  {Borne}, {Bradac}, {Brandt}, {Bridge}, {Brown}, {Brunner}, {Bullock},
  {Burgasser}, {Burge}, {Burke}, {Cargile}, {Chandrasekharan}, {Chartas},
  {Chesley}, {Chu}, {Cinabro}, {Claire}, {Claver}, {Clowe}, {Connolly}, {Cook},
  {Cooke}, {Cooray}, {Covey}, {Culliton}, {de Jong}, {de Vries}, {Debattista},
  {Delgado}, {Dell'Antonio}, {Dhital}, {Di Stefano}, {Dickinson}, {Dilday},
  {Djorgovski}, {Dobler}, {Donalek}, {Dubois-Felsmann}, {Durech},
  {Eliasdottir}, {Eracleous}, {Eyer}, {Falco}, {Fan}, {Fassnacht}, {Ferguson},
  {Fernandez}, {Fields}, {Finkbeiner}, {Figueroa}, {Fox}, {Francke}, {Frank},
  {Frieman}, {Fromenteau}, {Furqan}, {Galaz}, {Gal-Yam}, {Garnavich},
  {Gawiser}, {Geary}, {Gee}, {Gibson}, {Gilmore}, {Grace}, {Green}, {Gressler},
  {Grillmair}, {Habib}, {Haggerty}, {Hamuy}, {Harris}, {Hawley}, {Heavens},
  {Hebb}, {Henry}, {Hileman}, {Hilton}, {Hoadley}, {Holberg}, {Holman},
  {Howell}, {Infante}, {Ivezic}, {Jacoby}, {Jain}, {R}, {Jedicke}, {Jee},
  {Garrett Jernigan}, {Jha}, {Johnston}, {Jones}, {Juric}, {Kaasalainen},
  {Styliani}, {Kafka}, {Kahn}, {Kaib}, {Kalirai}, {Kantor}, {Kasliwal},
  {Keeton}, {Kessler}, {Knezevic}, {Kowalski}, {Krabbendam}, {Krughoff},
  {Kulkarni}, {Kuhlman}, {Lacy}, {Lepine}, {Liang}, {Lien}, {Lira}, {Long},
  {Lorenz}, {Lotz}, {Lupton}, {Lutz}, {Macri}, {Mahabal}, {Mandelbaum},
  {Marshall}, {May}, {McGehee}, {Meadows}, {Meert}, {Milani}, {Miller},
  {Miller}, {Mills}, {Minniti}, {Monet}, {Mukadam}, {Nakar}, {Neill}, {Newman},
  {Nikolaev}, {Nordby}, {O'Connor}, {Oguri}, {Oliver}, {Olivier}, {Olsen},
  {Olsen}, {Olszewski}, {Oluseyi}, {Padilla}, {Parker}, {Pepper}, {Peterson},
  {Petry}, {Pinto}, {Pizagno}, {Popescu}, {Prsa}, {Radcka}, {Raddick},
  {Rasmussen}, {Rau}, {Rho}, {Rhoads}, {Richards}, {Ridgway}, {Robertson},
  {Roskar}, {Saha}, {Sarajedini}, {Scannapieco}, {Schalk}, {Schindler},
  {Schmidt}, {Schmidt}, {Schneider}, {Schumacher}, {Scranton}, {Sebag},
  {Seppala}, {Shemmer}, {Simon}, {Sivertz}, {Smith}, {Allyn Smith}, {Smith},
  {Spitz}, {Stanford}, {Stassun}, {Strader}, {Strauss}, {Stubbs}, {Sweeney},
  {Szalay}, {Szkody}, {Takada}, {Thorman}, {Trilling}, {Trimble}, {Tyson}, {Van
  Berg}, {Vanden Berk}, {VanderPlas}, {Verde}, {Vrsnak}, {Walkowicz},
  {Wandelt}, {Wang}, {Wang}, {Warner}, {Wechsler}, {West}, {Wiecha},
  {Williams}, {Willman}, {Wittman}, {Wolff}, {Wood-Vasey}, {Wozniak}, {Young},
  {Zentner}, \& {Zhan}}]{LSST2009}
{LSST Science Collaboration}, {Abell}, P.~A., {Allison}, J., {et~al.} 2009,
  arXiv e-prints, arXiv:0912.0201

\bibitem[{{Lucas} {et~al.}(2008){Lucas}, {Hoare}, {Longmore}, {Schr{\"o}der},
  {Davis}, {Adamson}, {Band yopadhyay}, {de Grijs}, {Smith}, {Gosling},
  {Mitchison}, {G{\'a}sp{\'a}r}, {Coe}, {Tamura}, {Parker}, {Irwin}, {Hambly},
  {Bryant}, {Collins}, {Cross}, {Evans}, {Gonzalez-Solares}, {Hodgkin},
  {Lewis}, {Read}, {Riello}, {Sutorius}, {Lawrence}, {Drew}, {Dye}, \&
  {Thompson}}]{Lucas2008}
{Lucas}, P.~W., {Hoare}, M.~G., {Longmore}, A., {et~al.} 2008, \mnras, 391, 136

\bibitem[{{Macri} {et~al.}(2019){Macri}, {Kraan-Korteweg}, {Lambert}, {Alonso},
  {Berlind}, {Calkins}, {Erdo{\u{g}}du}, {Falco}, {Jarrett}, \&
  {Mink}}]{Macri2019}
{Macri}, L.~M., {Kraan-Korteweg}, R.~C., {Lambert}, T., {et~al.} 2019, \apjs,
  245, 6

\bibitem[{{Massaro} {et~al.}(2012){Massaro}, {D'Abrusco}, {Tosti}, {Ajello},
  {Gasparrini}, {Grindlay}, \& {Smith}}]{Massaro2012}
{Massaro}, F., {D'Abrusco}, R., {Tosti}, G., {et~al.} 2012, \apj, 750, 138

\bibitem[{{Minniti} {et~al.}(2010){Minniti}, {Lucas}, {Emerson}, {Saito},
  {Hempel}, {Pietrukowicz}, {Ahumada}, {Alonso}, {Alonso-Garcia}, {Arias},
  {Bandyopadhyay}, {Barb{\'a}}, {Barbuy}, {Bedin}, {Bica}, {Borissova},
  {Bronfman}, {Carraro}, {Catelan}, {Clari{\'a}}, {Cross}, {de Grijs},
  {D{\'e}k{\'a}ny}, {Drew}, {Fari{\~n}a}, {Feinstein}, {Fern{\'a}ndez
  Laj{\'u}s}, {Gamen}, {Geisler}, {Gieren}, {Goldman}, {Gonzalez}, {Gunthardt},
  {Gurovich}, {Hambly}, {Irwin}, {Ivanov}, {Jord{\'a}n}, {Kerins}, {Kinemuchi},
  {Kurtev}, {L{\'o}pez-Corredoira}, {Maccarone}, {Masetti}, {Merlo},
  {Messineo}, {Mirabel}, {Monaco}, {Morelli}, {Padilla}, {Palma}, {Parisi},
  {Pignata}, {Rejkuba}, {Roman-Lopes}, {Sale}, {Schreiber}, {Schr{\"o}der},
  {Smith}, {}, {Soto}, {Tamura}, {Tappert}, {Thompson}, {Toledo}, {Zoccali}, \&
  {Pietrzynski}}]{Minniti2010}
{Minniti}, D., {Lucas}, P.~W., {Emerson}, J.~P., {et~al.} 2010, \na, 15, 433

\bibitem[{{Minniti} {et~al.}(2018){Minniti}, {Saito}, {Gonzalez},
  {Alonso-Garc{\'\i}a}, {Rejkuba}, {Barb{\'a}}, {Irwin}, {Kammers}, {Lucas},
  {Majaess}, \& {Valenti}}]{Minniti2018}
{Minniti}, D., {Saito}, R.~K., {Gonzalez}, O.~A., {et~al.} 2018, \aap, 616, A26

\bibitem[{{Paladini} {et~al.}(2023){Paladini}, {Zucker}, {Benjamin}, {Nataf},
  {Minniti}, {Zasowski}, {Peek}, {Carey}, {Allen}, {Alonso-Garcia}, {Alves},
  {Anders}, {Athanassoula}, {Beers}, {Bird}, {Bland-Hwathorn}, {Brown},
  {Buder}, {Casagrande}, {Casey}, {Cassisi}, {Catelan}, {Chary}, {Chene},
  {Ciardi}, {Comeron}, {Cohen}, {Dame}, {Drimmel}, {Fernandez Trincado},
  {Finkbeiner}, {Geisler}, {Gennaro}, {Goodman}, {Green}, {Hajdu}, {Henderson},
  {Hora}, {Ivanov}, {Kirkpatrick}, {Kobayashi}, {Kuhn}, {Kunder}, {Lu},
  {Lucas}, {Majaess}, {Megeath}, {Meisner}, {Molinari}, {Mroz}, {Ness},
  {Neumayer}, {Nogueras-Lara}, {Noriega-Crespo}, {Poleski}, {Rix}, {Rebull},
  {Reggiani}, {Rejkuba}, {Saito}, {Schoenrich}, {Saydjari}, {Schisano},
  {Schlafly}, {Schlaufman}, {Smith}, {Speagle}, {Wisz}, {Wyse}, \&
  {Zakamska}}]{Paladini2023}
{Paladini}, R., {Zucker}, C., {Benjamin}, R., {et~al.} 2023, arXiv e-prints,
  arXiv:2307.07642

\bibitem[{{Pichel} {et~al.}(2020){Pichel}, {Donoso}, {Baravalle}, {Alonso},
  {Rovero}, {Beam{\'\i}n}, {Minniti}, {Cabral}, {S{\'a}nchez}, {Coldwell}, \&
  {Masetti}}]{Pichel2020}
{Pichel}, A., {Donoso}, L.~G., {Baravalle}, L.~D., {et~al.} 2020, \mnras, 491,
  3448

\bibitem[{{Quintana} {et~al.}(1995){Quintana}, {Ramirez}, {Melnick},
  {Raychaudhury}, \& {Slezak}}]{Quintana1995}
{Quintana}, H., {Ramirez}, A., {Melnick}, J., {Raychaudhury}, S., \& {Slezak},
  E. 1995, \aj, 110, 463

\bibitem[{{Radburn-Smith} {et~al.}(2006){Radburn-Smith}, {Lucey}, {Woudt},
  {Kraan-Korteweg}, \& {Watson}}]{Radburn2006}
{Radburn-Smith}, D.~J., {Lucey}, J.~R., {Woudt}, P.~A., {Kraan-Korteweg},
  R.~C., \& {Watson}, F.~G. 2006, \mnras, 369, 1131

\bibitem[{{Rajohnson} {et~al.}(2024{\natexlab{a}}){Rajohnson},
  {Kraan-Korteweg}, {Chen}, {Frank}, {Steyn}, {Kurapati}, {Pisano},
  {Staveley-Smith}, {Serra}, {Goedhart}, \& {Camilo}}]{Rajohnson2024}
{Rajohnson}, S. H.~A., {Kraan-Korteweg}, R.~C., {Chen}, H., {et~al.}
  2024{\natexlab{a}}, \mnras, 531, 3486

\bibitem[{{Rajohnson} {et~al.}(2024{\natexlab{b}}){Rajohnson},
  {Kraan-Korteweg}, {Frank}, {Chen}, {Staveley-Smith}, {Serra}, {Steyn},
  {Kurapati}, {Pisano}, \& {Goedhart}}]{2024MNRAS.535.3429R}
{Rajohnson}, S. H.~A., {Kraan-Korteweg}, R.~C., {Frank}, B.~S., {et~al.}
  2024{\natexlab{b}}, \mnras, 535, 3429

\bibitem[{{Ramatsoku} {et~al.}(2016){Ramatsoku}, {Verheijen}, {Kraan-Korteweg},
  {J{\'o}zsa}, {Schr{\"o}der}, {Jarrett}, {Elson}, {van Driel}, {de Blok}, \&
  {Henning}}]{Ramatsoku2016}
{Ramatsoku}, M., {Verheijen}, M.~A.~W., {Kraan-Korteweg}, R.~C., {et~al.} 2016,
  Mon. Not. R. Astron. Soc., 460, 923

\bibitem[{{Saito} {et~al.}(2024){Saito}, Hempel, Alonso-García, Lucas,
  Minniti, Alonso, Baravalle, Borissova, Caceres, Chené, Cross, Duplancic,
  Garro, Gómez, Ivanov, Kurtev, Luna, Majaess, Navarro, Pullen, Rejkuba,
  Sanders, Smith, Albino, Alonso, Amôres, Angeloni, Arias, Arnaboldi, Barbuy,
  Bayo, Beamin, Bedin, Bellini, Benjamin, Bica, Bonatto, Botan, Braga, Brown,
  Cabral, Camargo, Garatti, Carballo-Bello, Catelan, Chavero, Chijani, Clariá,
  Coldwell, Contreras~Peña, Contreras~Ramos, Corral-Santana, Cortés,
  Cortés-Contreras, Cruz, Daza-Perilla, Debattista, Dias, Donoso, D'Souza,
  Emerson, Federle, Fermiano, Fernandez, Fernández-Trincado, Ferreira,
  Ferreira~Lopes, Firpo, Flores-Quintana, Fraga, Froebrich, Galdeano,
  Gavignaud, Geisler, Gerhard, Gieren, Gonzalez, Gramajo, Gran, Granitto,
  Griggio, Guo, Gurovich, Hilker, Jones, Kammers, Kuhn, Kumar, Kundu, Lares,
  Libralato, Lima, Maccarone, Marchant~Cortés, Martin, Masetti, Matsunaga,
  Mauro, McDonald, Mejías, Mesa, Milla-Castro, Minniti, Moni~Bidin,
  Montenegro, Morris, Motta, Navarete, Navarro~Molina, Nikzat, Nilo~Castellón,
  Obasi, Ortigoza-Urdaneta, Palma, Parisi, Pena~Ramírez, Pereyra, Perez,
  Petralia, Pichel, Pignata, Ramírez~Alegría, Rojas, Rojas, Roman-Lopes,
  Rovero, Saroon, Schmidt, Schröder, Schultheis, Sgró, Solano, Soto,
  Stecklum, Steeghs, Tamura, Tissera, Valcarce, Valotto, Vasquez, Villalon,
  Villanova, Vivanco~Cádiz, Zelada~Bacigalupo, Zijlstra, \&
  Zoccali}]{Saito2024}
{Saito}, R.~K., Hempel, M., Alonso-García, J., {et~al.} 2024, \aap, 689, 19

\bibitem[{{Saito} {et~al.}(2012){Saito}, {Minniti}, {Dias}, {Hempel},
  {Rejkuba}, {Alonso-Garc{\'{\i}}a}, {Barbuy}, {Catelan}, {Emerson},
  {Gonzalez}, {Lucas}, \& {Zoccali}}]{Saito2012}
{Saito}, R.~K., {Minniti}, D., {Dias}, B., {et~al.} 2012, \aap, 544, A147

\bibitem[{{Schlafly} \& {Finkbeiner}(2011)}]{Schlafly2011}
{Schlafly}, E.~F. \& {Finkbeiner}, D.~P. 2011, \apj, 737, 103

\bibitem[{{Schr{\"o}der} {et~al.}(2009){Schr{\"o}der}, {Kraan-Korteweg}, \&
  {Henning}}]{Schroder2009}
{Schr{\"o}der}, A.~C., {Kraan-Korteweg}, R.~C., \& {Henning}, P.~A. 2009, \aap,
  505, 1049

\bibitem[{{Sersic}(1968)}]{Sersic1968}
{Sersic}, J.~L. 1968, {Atlas de galaxias australes}

\bibitem[{Shapley(1961)}]{Shapley1961}
Shapley, H. 1961, Journal of the Royal Astronomical Society of Canada, 55, 273

\bibitem[{{Sharples} {et~al.}(2013){Sharples}, {Bender}, {Agudo Berbel},
  {Bezawada}, {Castillo}, {Cirasuolo}, {Davidson}, {Davies}, {Dubbeldam},
  {Fairley}, {Finger}, {F{\"o}rster Schreiber}, {Gonte}, {Hess}, {Jung},
  {Lewis}, {Lizon}, {Muschielok}, {Pasquini}, {Pirard}, {Popovic}, {Ramsay},
  {Rees}, {Richter}, {Riquelme}, {Rodrigues}, {Saviane}, {Schlichter},
  {Schmidtobreick}, {Segovia}, {Smette}, {Szeifert}, {van Kesteren}, {Wegner},
  \& {Wiezorrek}}]{Sharples2013}
{Sharples}, R., {Bender}, R., {Agudo Berbel}, A., {et~al.} 2013, The Messenger,
  151, 21

\bibitem[{{Skelton} {et~al.}(2009){Skelton}, {Woudt}, \&
  {Kraan-Korteweg}}]{Skelton2009}
{Skelton}, R.~E., {Woudt}, P.~A., \& {Kraan-Korteweg}, R.~C. 2009, \mnras, 396,
  2367

\bibitem[{{Skrutskie} {et~al.}(2006){Skrutskie}, {Cutri}, {Stiening},
  {Weinberg}, {Schneider}, {Carpenter}, {Beichman}, {Capps}, {Chester},
  {Elias}, {Huchra}, {Liebert}, {Lonsdale}, {Monet}, {Price}, {Seitzer},
  {Jarrett}, {Kirkpatrick}, {Gizis}, {Howard}, {Evans}, {Fowler}, {Fullmer},
  {Hurt}, {Light}, {Kopan}, {Marsh}, {McCallon}, {Tam}, {Van Dyk}, \&
  {Wheelock}}]{Skrutskie2006}
{Skrutskie}, M.~F., {Cutri}, R.~M., {Stiening}, R., {et~al.} 2006, \aj, 131,
  1163

\bibitem[{{Soto} {et~al.}(2013){Soto}, {Barb{\'a}}, {Gunthardt}, {Minniti},
  {Lucas}, {Majaess}, {Irwin}, {Emerson}, {Gonzalez-Solares}, {Hempel},
  {Saito}, {Gurovich}, {Roman-Lopes}, {Moni-Bidin}, {Santucho}, {Borissova},
  {Kurtev}, {Toledo}, {Geisler}, {Dominguez}, \& {Beamin}}]{Soto2013}
{Soto}, M., {Barb{\'a}}, R., {Gunthardt}, G., {et~al.} 2013, \aap, 552, A101

\bibitem[{{Staveley-Smith} {et~al.}(2016){Staveley-Smith}, {Kraan-Korteweg},
  {Schr{\"o}der}, {Henning}, {Koribalski}, {Stewart}, \&
  {Heald}}]{Staveley2016}
{Staveley-Smith}, L., {Kraan-Korteweg}, R.~C., {Schr{\"o}der}, A.~C., {et~al.}
  2016, Astron. J., 151, 52

\bibitem[{{Steyn} {et~al.}(2024){Steyn}, {Kraan-Korteweg}, {Rajohnson},
  {Kurapati}, {Chen}, {Frank}, {Serra}, {Staveley-Smith}, {Camilo}, \&
  {Goedhart}}]{Steyn2024}
{Steyn}, N., {Kraan-Korteweg}, R.~C., {Rajohnson}, S. H.~A., {et~al.} 2024,
  \mnras, 529, L88

\bibitem[{{Su} {et~al.}(2013){Su}, {Kong}, {Li}, \& {Fang}}]{Shanshan2013}
{Su}, S., {Kong}, X., {Li}, J., \& {Fang}, G. 2013, \apj, 778, 10

\bibitem[{{Tully} {et~al.}(2014){Tully}, {Courtois}, {Hoffman}, \&
  {Pomar{\`e}de}}]{Tully2014}
{Tully}, R.~B., {Courtois}, H., {Hoffman}, Y., \& {Pomar{\`e}de}, D. 2014,
  Nature, 513, 71

\bibitem[{{Visvanathan} \& {Yamada}(1996)}]{Visvanathan1996}
{Visvanathan}, N. \& {Yamada}, T. 1996, \apjs, 107, 521

\bibitem[{{Westmeier} {et~al.}(2022){Westmeier}, {Deg}, {Spekkens}, {Reynolds},
  {Shen}, {Gaudet}, {Goliath}, {Huynh}, {Venkataraman}, {Lin}, {O'Beirne},
  {Catinella}, {Cortese}, {D{\'e}nes}, {Elagali}, {For}, {J{\'o}zsa},
  {Howlett}, {van der Hulst}, {Jurek}, {Kamphuis}, {Kilborn}, {Kleiner},
  {Koribalski}, {Lee-Waddell}, {Murugeshan}, {Rhee}, {Serra}, {Shao},
  {Staveley-Smith}, {Wang}, {Wong}, {Zwaan}, {Allison}, {Anderson}, {Ball},
  {Bock}, {Brodrick}, {Bunton}, {Cooray}, {Gupta}, {Hayman}, {Mahony}, {Moss},
  {Ng}, {Pearce}, {Raja}, {Roxby}, {Voronkov}, {Warhurst}, {Courtois}, \&
  {Said}}]{Westmeier2022}
{Westmeier}, T., {Deg}, N., {Spekkens}, K., {et~al.} 2022, \pasa, 39, e058

\bibitem[{{Williams} {et~al.}(2014){Williams}, {Kraan-Korteweg}, \&
  {Woudt}}]{Williams2014}
{Williams}, W.~L., {Kraan-Korteweg}, R.~C., \& {Woudt}, P.~A. 2014, \mnras,
  443, 41

\bibitem[{{Woudt}(1998)}]{Woudt1998}
{Woudt}, P.~A. 1998, PhD thesis, University of Cape Town, South Africa

\bibitem[{{Woudt} \& {Kraan-Korteweg}(2001)}]{Woudt2001}
{Woudt}, P.~A. \& {Kraan-Korteweg}, R.~C. 2001, \aap, 380, 441

\bibitem[{{Wright} {et~al.}(2010){Wright}, {Eisenhardt}, {Mainzer}, {Ressler},
  {Cutri}, {Jarrett}, {Kirkpatrick}, {Padgett}, {McMillan}, {Skrutskie},
  {Stanford}, {Cohen}, {Walker}, {Mather}, {Leisawitz}, {Gautier}, {McLean},
  {Benford}, {Lonsdale}, {Blain}, {Mendez}, {Irace}, {Duval}, {Liu}, {Royer},
  {Heinrichsen}, {Howard}, {Shannon}, {Kendall}, {Walsh}, {Larsen}, {Cardon},
  {Schick}, {Schwalm}, {Abid}, {Fabinsky}, {Naes}, \& {Tsai}}]{Wright2010}
{Wright}, E.~L., {Eisenhardt}, P.~R.~M., {Mainzer}, A.~K., {et~al.} 2010, \aj,
  140, 1868

\bibitem[{{Zurita Heras} {et~al.}(2009){Zurita Heras}, {Chaty}, \&
  {Tomsick}}]{Zurita2009}
{Zurita Heras}, J.~A., {Chaty}, S., \& {Tomsick}, J.~A. 2009, \aap, 502, 787

\end{thebibliography}

\end{document}